\documentclass[]{aa}

\begin{document}
\title{Chromospheric evaporation and phase mixing of Alfv\'en waves in coronal loops}

\author{H.J.~Van Damme\inst{\ref{inst1}} \and
          I.~De Moortel\inst{\ref{inst1},\ref{inst2}} \and
          P.~Pagano\inst{\ref{inst1}} \and
          C.D.~Johnston\inst{\ref{inst1}}
          }

\authorrunning{Van Damme et al.}
\titlerunning{Evaporation due to Alfv\'en wave phase mixing}

\institute{School of Mathematics and Statistics, University of St Andrews, North Haugh, St Andrews, KY16 9SS, UK \label{inst1}
            \and
            Rosseland Centre for Solar Physics, University of Oslo, PO Box 1029 Blindern, NO-0315 Oslo, Norway \label{inst2}
 \\
      \email{hjvd@st-andrews.ac.uk}
      }

   \date{ }

\abstract{Phase mixing of Alfv\'en waves has been studied extensively as a possible coronal heating mechanism but without the full thermodynamic consequences considered self-consistently. It has been argued that in some cases, the thermodynamic feedback of the heating could substantially affect the transverse density gradient and even inhibit the phase mixing process.}
{In this paper, for the first time, we use MHD simulations with the appropriate thermodynamical terms included to quantify the evaporation following heating by phase mixing of Alfv\'en waves in a coronal loop and the effect of this evaporation on the transverse density profile.}
{The numerical simulations were performed using the Lagrangian Remap code Lare2D. We set up a 2D loop model consisting of a field-aligned thermodynamic equilibrium and a cross-field (background) heating profile. A continuous, sinusoidal, high-frequency Alfv\'en wave driver was implemented. As the Alfv\'en waves propagate along the field, they undergo phase mixing due to the cross-field density gradient in the coronal part of the loop. We investigated the presence of field-aligned flows, heating from the dissipation of the phase-mixed Alfv\'en waves, and the subsequent evaporation from the lower atmosphere.}
{We find that phase mixing of Alfv\'en waves leads to modest heating in the shell regions of the loop and evaporation of chromospheric material into the corona with upflows of the order of only 5-20 m/s. Although the evaporation leads to a mass increase in the shell regions of the loop, the effect on the density gradient and, hence, on the phase mixing process, is insignificant.}
{This paper self-consistently investigates the effect of chromospheric evaporation on the cross-field density gradient and the phase mixing process in a coronal loop. We found that the effects in our particular setup (small amplitude, high frequency waves) are too small to significantly change the density gradient. }

\maketitle


\section{Introduction}\label{sec:introduction}

Recent high-cadence and high-resolution observations have established the presence of waves and oscillations throughout the solar atmosphere (see e.g. reviews by \citealt{Nakariakov2005}, \citealt{Banerjee2007}; \citealt{Zaq2009}; \citealt{Arregui2012}; \citealt{paper:DeMoortel2012}, \citealt{Mathi2013}, \citealt{paper:Arregui2015}, \citealt{Jess2015}).
Most of these perturbations have been interpreted as MHD waves and in many cases are reported to contain a substantial amount of energy, leading to a renewed interest in MHD wave dissipation as a potential coronal heating mechanism \citep[e.g.][]{paper:ParnellDeMoortel2012, paper:Arregui2015}. 

Phase mixing of Alfv\'en waves \citep{paper:Priest1983} is one of the mechanisms proposed to address the slow dissipation of wave energy with classical transport coefficients; in the presence of a cross-field gradient in the Alfv\'en speed, waves on neighbouring field lines move out of phase, building up large gradients which, in turn, lead to enhanced dissipation. 

\begin{figure*}
\centering
\includegraphics[width=0.35\textwidth]{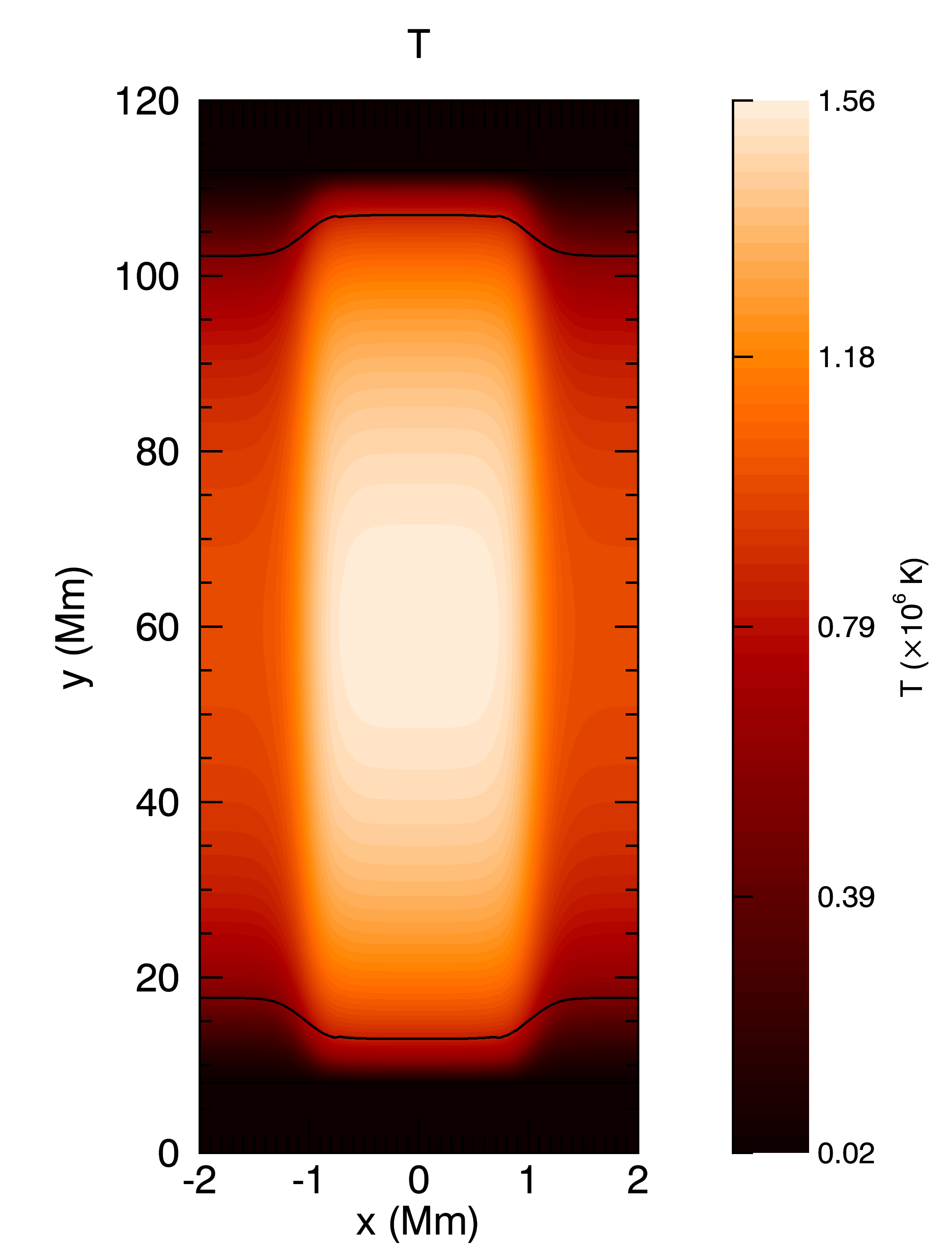}
\includegraphics[width=0.35\textwidth]{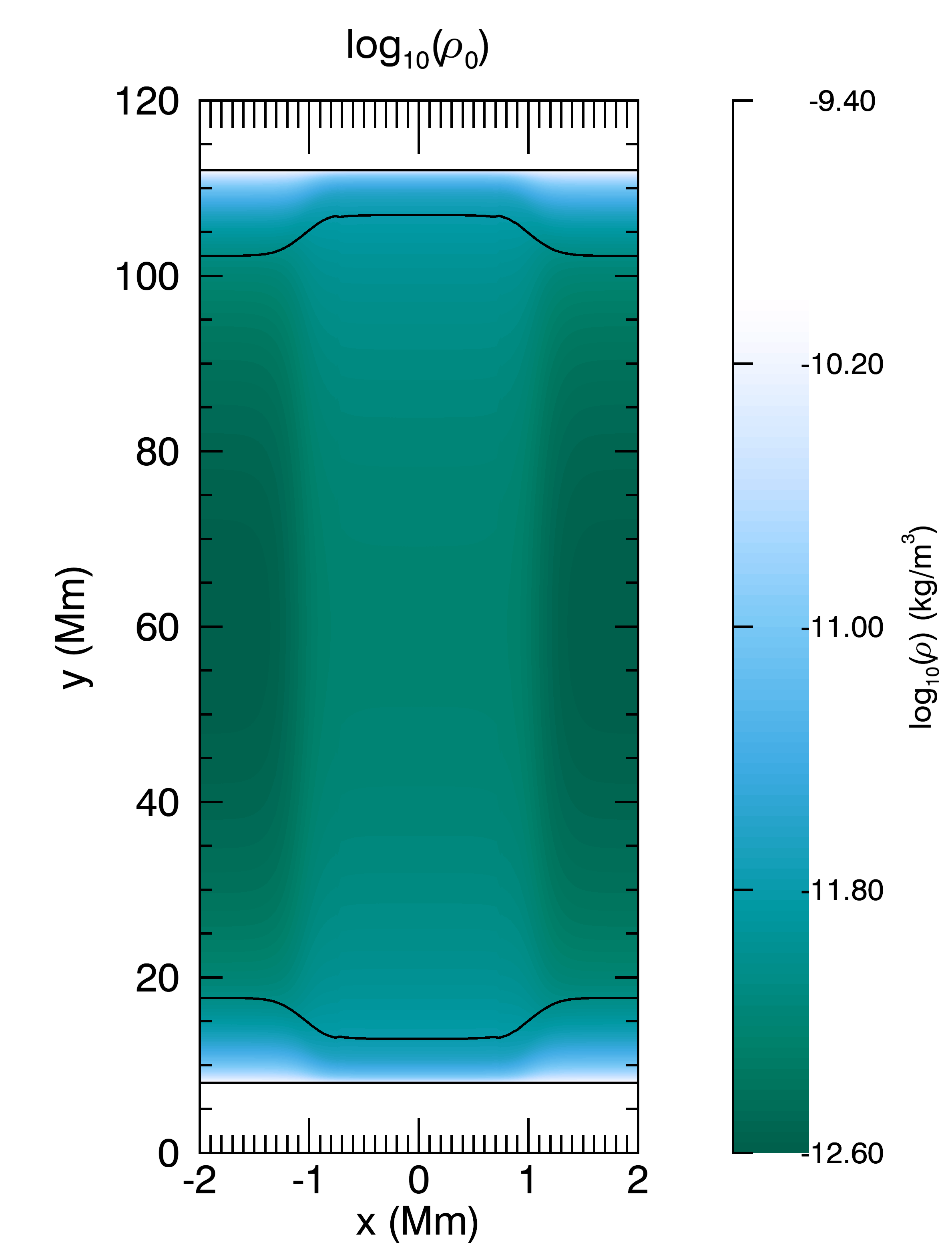}
    \caption{Contour plots of the temperature (left) and density (right) after the numerical relaxation.}
\label{fig_init_cond1}
\end{figure*}

As heating occurs in the corona, an enhanced conductive flux is driven downward from the corona to the lower atmosphere which, in turn, leads to an upward enthalpy flux (evaporation), increasing the coronal density. Unless heating continues, the corona will then start to cool whilst the density continues to increase until a balance between thermal conduction and radiation is reached.  In the final stage, as the corona continues to cool, mass starts to drain from the loop. \citep[see e.g.][]{paper:Cargill1994, paper:Klimchuk2006}. \cite{Hansteen2010}, for example, studied this mass and energy cycle between the lower atmosphere and the corona and found that both downflows and upflows are present at locations of (strong) magnetic field braiding, leading to redshifts and blueshifts of the order of about 5 km/s \citep[see also e.g.][]{Zacharias2011, Guerreiro2013}. Observationally, a multitude of studies have found evidence for the presence of a complex interplay between upflows and downflows in the Transition Region (TR) and lower corona \citep[e.g.][]{DelZanna2008, Feldman2011, Dadashi2011, Dadashi2012, Tripathi2012, TripathiMason2012, McIntosh2012, Winebarger2013}. A comprehensive review of the modelling of impulsive heating in coronal loops can be found in e.g. \cite{Reale2014}. \cite{paper:Bradshaw2013} presented a detailed discussion on the importance of numerical resolution in the TR to accurately model the mass and energy exchange. Various solutions have been proposed to address this problem such as (controlled) broadening of the TR \citep[e.g.][]{paper:Lionello2009,paper:Mikic2013, JohnstonBrad2019, Johnston2020} and a jump-condition approach \citep{paper:Johnston2017a, paper:Johnston2017b,Johnston2019}.

Although phase mixing of Alfv\'en waves has been studied extensively as a potential coronal heating mechanism, it has usually been investigated without taking into account the mass exchange with the lower atmosphere described above. In particular, heating could change the local density profile (e.g.~through evaporation or draining) and, hence, the Alfv\'en speed profile which, in turn, could alter the phase mixing process. This feedback mechanism was investigated for resonant absorption by \cite{Ofman1998}. These authors used thermodynamic equilibrium scaling laws to instantaneously adjust the local plasma parameters and found that the feedback caused the heating layers to drift. However, \cite{paper:Cargill2016} show that the `detuning' of the local resonance conditions is a relatively slow process and hence that the instantaneous update to the local plasma conditions used by \cite{Ofman1998} is likely to overestimate the effect of the feedback mechanism. 

\cite{paper:Cargill2016} also demonstrated conceptually that heating by phase mixing of Alfv\'en waves cannot sustain the (assumed) background density profile as heating takes place preferentially where the Alfv\'en speed (or density) gradient is largest, but cooling and draining occur fastest where the density is higher. In other words, the spatial profile of the heating by phase mixing is not compatible with the imposed density structuring. These authors also found that although the heating feedback mechanism could lead to a highly structured density profile, this only happened on long timescales, comparable to the loop cooling and draining timescales and, hence, the changes in the density profile are dominated by the thermal evolution.

In this paper, we present 2D MHD simulations of phase mixing of Alfv\'en waves in a coronal loop, including optically thin radiation and thermal conduction, and we study the plasma heating and subsequent evaporation of mass from the lower atmosphere.
This allows us to investigate whether the dissipation of phase-mixed Alfv\'en waves does indeed lead to evaporation from the lower atmosphere and whether such mass flows in turn affect the phase mixing process. This paper is structured as follows. In Section \ref{s:Model}, we introduce the methods and numerical setup for the coronal loop model used in this study. The results are presented in Section \ref{s:Results}, followed by the discussion and conclusions in Section \ref{s:Discussion}. 


\section{Model and numerical method} \label{s:Model}

\subsection{Equations and initial conditions}

We use the 2D Lagrangian Remap code Lare2D \citep{paper:Arber2001} to solve the (normalised) MHD equations:
\begin{align}
 \frac{\partial\rho}{\partial t}&=-\nabla\cdot\left(\rho \mathbf{v}\right), \label{cont_eq}\\
 \rho\frac{D\mathbf{v}}{Dt}&=-\nabla P-\rho \mathbf{g} +\mathbf{j}\times \mathbf{B}+ \rho\nu\left(\nabla^2\mathbf{v}\right),\label{momentum_eq} \\
 \rho\frac{D\epsilon}{Dt}&=-P\nabla\cdot \mathbf{v} - \nabla\cdot \mathbf{F_{c}}- \rho^2\Lambda(T)+H_{bg}(x)\\
 &+\eta j^2 + Q_{visc}, \label{energy_eq}\\
 P&=2\rho T,\label{ideal_gas_law}\\
 \frac{\partial \bf{B}}{\partial t} &= \nabla\times\left(\mathbf{v}\times\mathbf{B}\right)- \nabla\times\left(\eta\nabla\times \mathbf{B}\right),\label{induction_eq} \\
\mathbf{j} &= \nabla\times \mathbf{B}. \label{Amperes_law}
\end{align}
Here, $D/Dt$ is the convective derivative, $\rho$ is the mass density, $\mathbf{v}$ the velocity, $P$ the gas pressure, $\mathbf{B}$ the magnetic field, $\mathbf{j}$ the current density, $\eta$ the resistivity, $\rho\nu$ the dynamic viscosity and $\epsilon =\frac{P}{(\gamma-1)\rho}$ the specific internal energy. We include a field-aligned gravitational acceleration $g_{||}$, where we use a semi-circular loop profile \citep[e.g.][]{paper:Reale2010}. 
Equation \eqref{energy_eq} is the energy equation where we include thermal conduction, optically thin radiation, a background heating function and the viscous heating term, $Q_{visc}=\rho\nu \nabla \mathbf{v}:(\nabla \mathbf{v})^T$. The conductive flux is given by $
\mathbf{F_c}=-\kappa_0 T^{5/2}(\mathbf{B}\cdot\nabla T)\mathbf{B}/B^2$
 and $\Lambda(T)=\chi T^\alpha$ is the optically thin radiative loss function, where $\chi$ and $\alpha$ are described in \cite{paper:Klimchuk2008}. 

To construct a 2D loop model with a cross-field density gradient, we combine a field-aligned hydrostatic (thermodynamic) equilibrium \citep[e.g.][]{paper:Reale2010,paper:Johnston2017a} with a background heating function that has a transverse (cross-field) profile. In our 2D Cartesian reference frame, $y$ is the field-aligned direction, $x$ the cross-field direction and $z$ is the invariant direction.
The non-uniform profile for the background heating function is given by
 \begin{equation}
 H_{bg}(x)=\frac{H_{1}+H_{2}}{2}+\frac{H_{2}-H_{1}}{2}\tanh(a(-|x|+1)),  \label{heating_profile}
\end{equation}
where $a=5$ , $H_2= 4 H_1$ and $H_{1}=3.24 \times 10^{-6}$ J m$^{-3}$s$^{-1}$ is the (background) heating required for a field-aligned (1D) equilibrium in the external region. After (numerical) relaxation, this combination of parameters results in a density ratio of $\rho_i/\rho_e =2.4$ at the loop apex, where $\rho_i$ and $\rho_e$ refer to the interior and exterior densities, respectively. The transverse profile of the heating function in Eq.~\ref{heating_profile} has been chosen as a model-representation of a concentrated background heating. Because of the transverse structuring of the heating function, a transverse density profile is created in the corona (a coronal `loop’) which will lead to phase mixing.

The numerical domain is 4 Mm $\times$ 120 Mm and consists of 256 gridpoints in $x$ (cross-field direction) and 4096 gridpoints in $y$ (field-aligned direction).  The boundary conditions are periodic in $x$ and zero gradient in $y$ with the velocity set to zero. The setup includes an isothermal model-chromosphere ($T = 2 \times 10^4$ K), at both ends of the loop, extending over the first and last 8 Mm of the domain. This model-chromosphere acts as a simple mass reservoir and does not include detailed chromospheric physics such as partial ionisation.

In order to fully resolve the TR, we use the approach proposed in \cite{paper:Lionello2009} and \cite{paper:Mikic2013}: below a fixed cut-off temperature $T_c$, the optically thin radiative loss rate is decreased and the parallel thermal conductivity is increased:
\begin{equation}
      \kappa_{||}(T) =\begin{cases} 
      \kappa_0 T^{5/2} & T \ge T_c \\
      \kappa_0 T_c^{5/2} & T < T_c. 
   \end{cases} \label{Mikic_conduction}
   \end{equation}
and  
   \begin{equation}
      \Lambda(T) =\begin{cases} 
      \Lambda(T) & T \ge T_c \\
      \Lambda(T)\left(\frac{T}{T_c}\right)^{5/2} & T < T_c. 
   \end{cases} \label{Mikic_radiation}
\end{equation}

This approach increases the field-aligned temperature length scale $L_T=T/\left|\frac{\partial T}{\partial y}\right|$, thus broadening the TR. In the simulations presented in this paper, we have set $T_c=5 \times 10^5\,$K. 

\begin{figure*}
\includegraphics[width=0.35\textwidth]{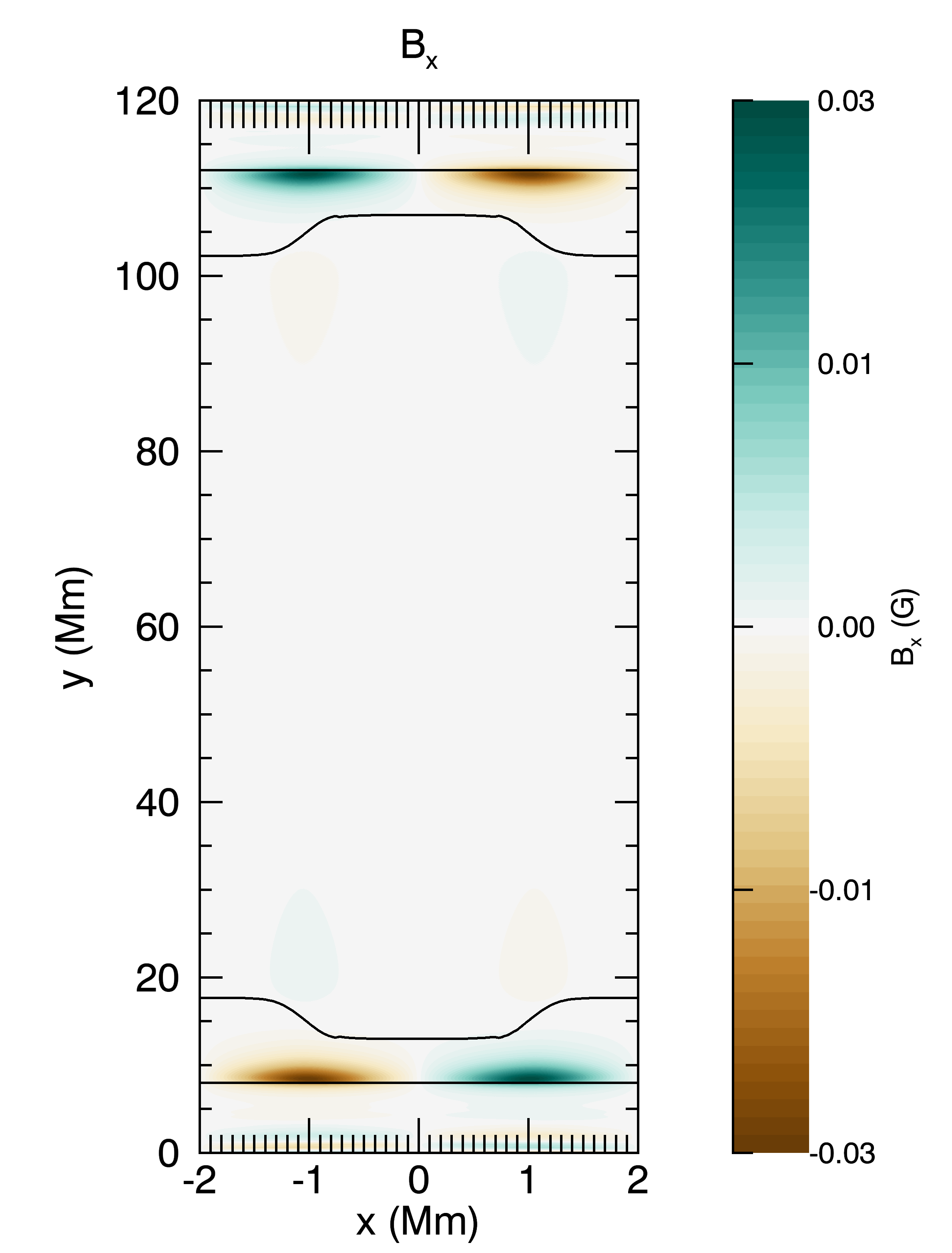}
\includegraphics[width=0.35\textwidth]{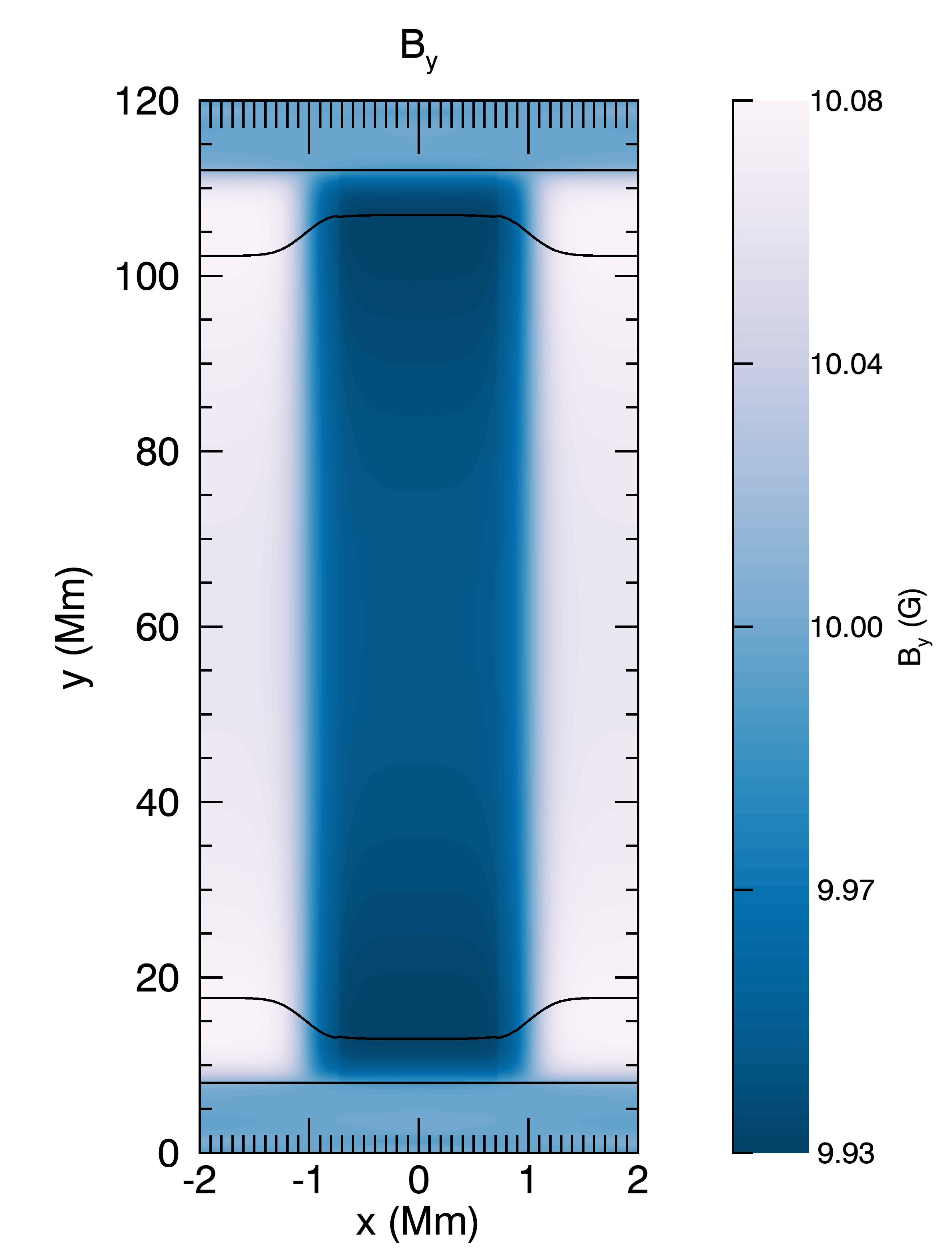}
\includegraphics[width=0.35\textwidth]{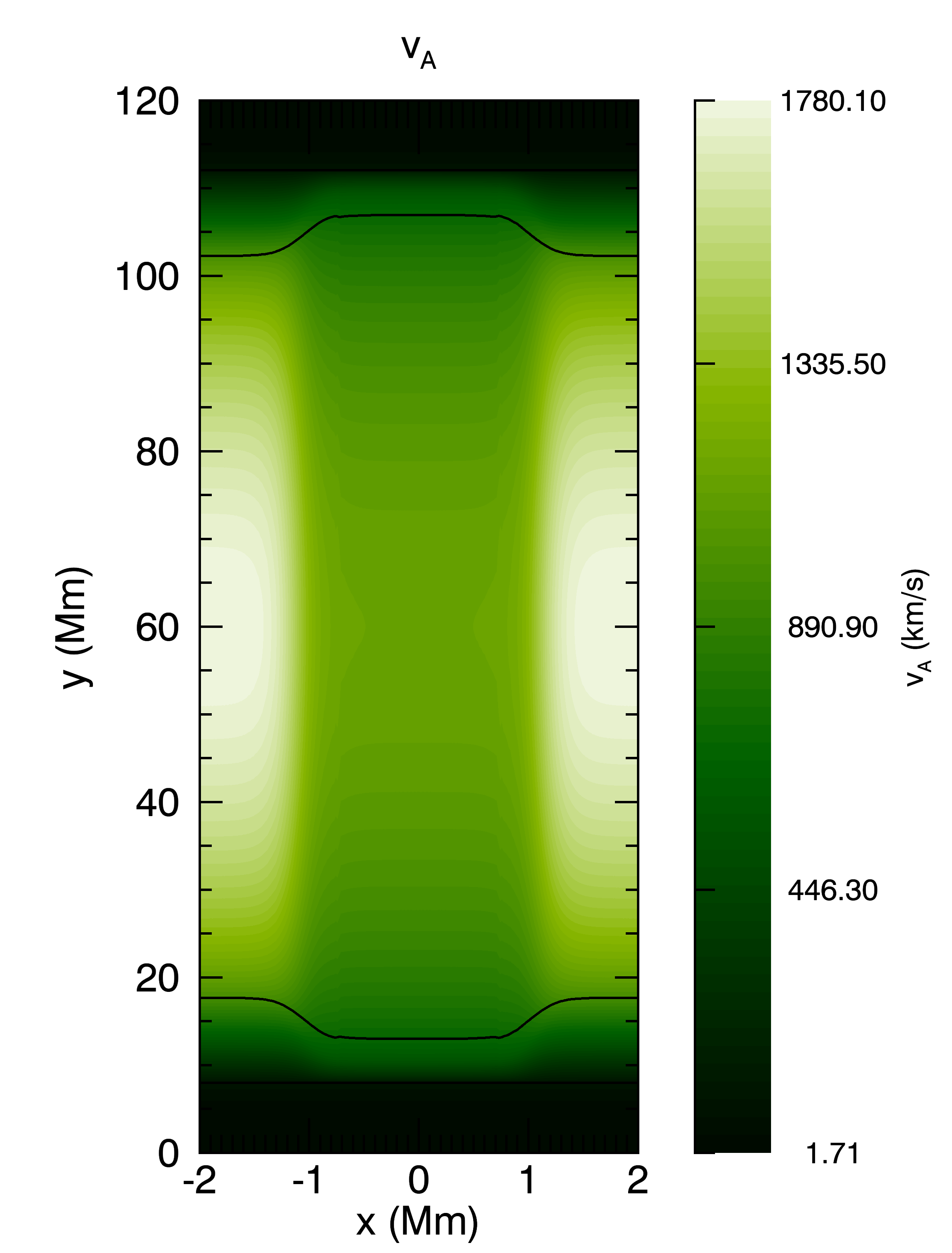}
\caption{Contour plots of the magnetic field components $B_x$ (left) and $B_y$ (middle) and the Alfv\'en speed (right) after the numerical relaxation.}
\label{fig_init_cond2}
\end{figure*}

Figure \ref{fig_init_cond1} shows contour plots of the temperature and density which are obtained after the numerical relaxation. In the coronal part of the loop, there is a clear cross-field gradient in the temperature and the density (as well as in the Alfv\'en speed, see Figure \ref{fig_init_cond2}), with the coronal temperature ranging between $1.0-1.6$ MK and the coronal density varying between $\rho=2.5-6 \times 10^{-13}$ kg/m$^{3}$ (giving a density ratio of $\rho_i / \rho_e = 2.4$ at the apex). This density range is lower than values usually measured in Active Region loops but can be seen as representative of the Quiet Sun \citep[see e.g.][]{Brooks2009,Brooks2012,Brooks2019}. The key aspect for phase mixing is the density contrast which, although modest, is reasonable for Quiet Sun loops.

\begin{figure*}
  \centering
   \includegraphics[width=0.45\textwidth]{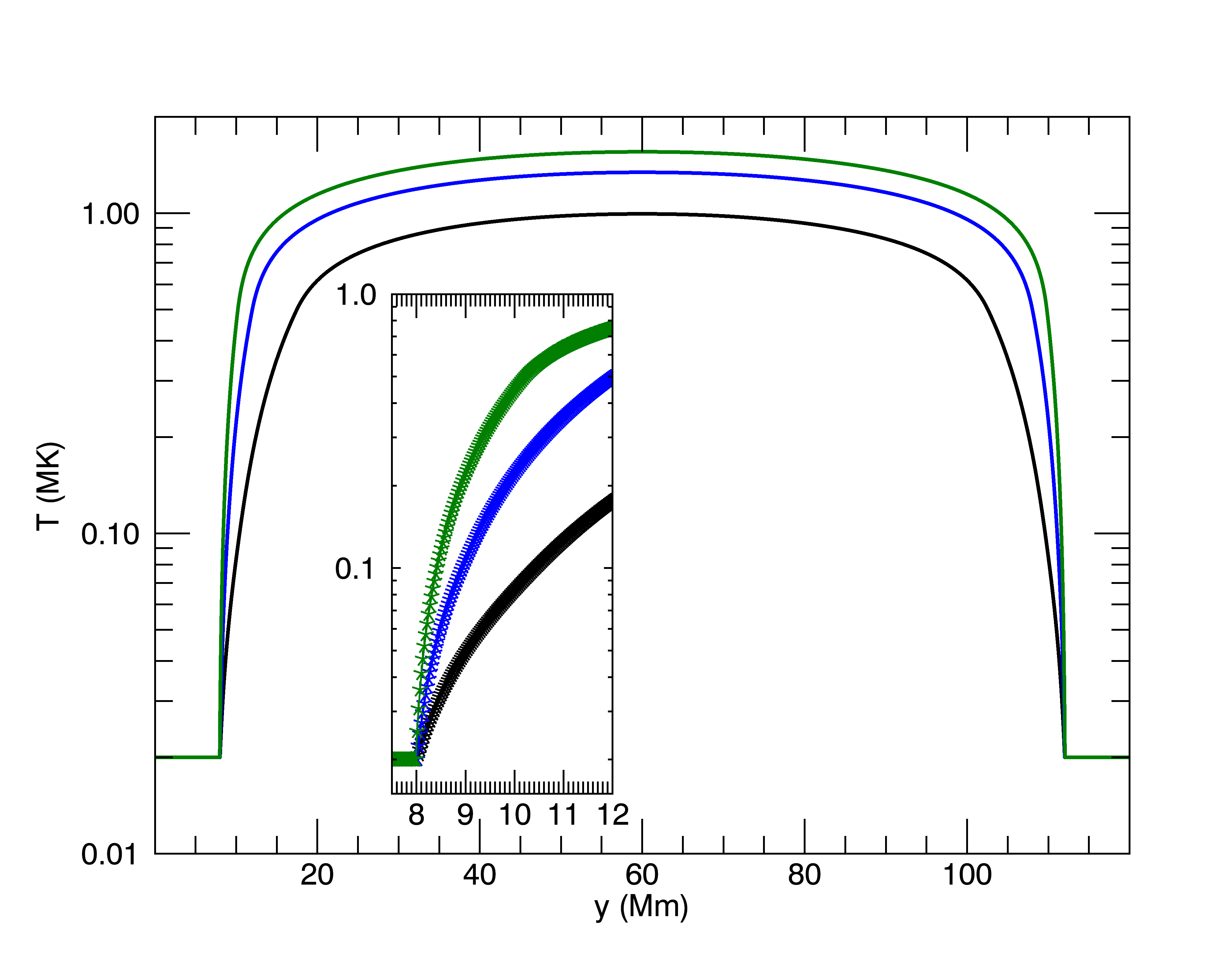}
   \includegraphics[width=0.45\textwidth]{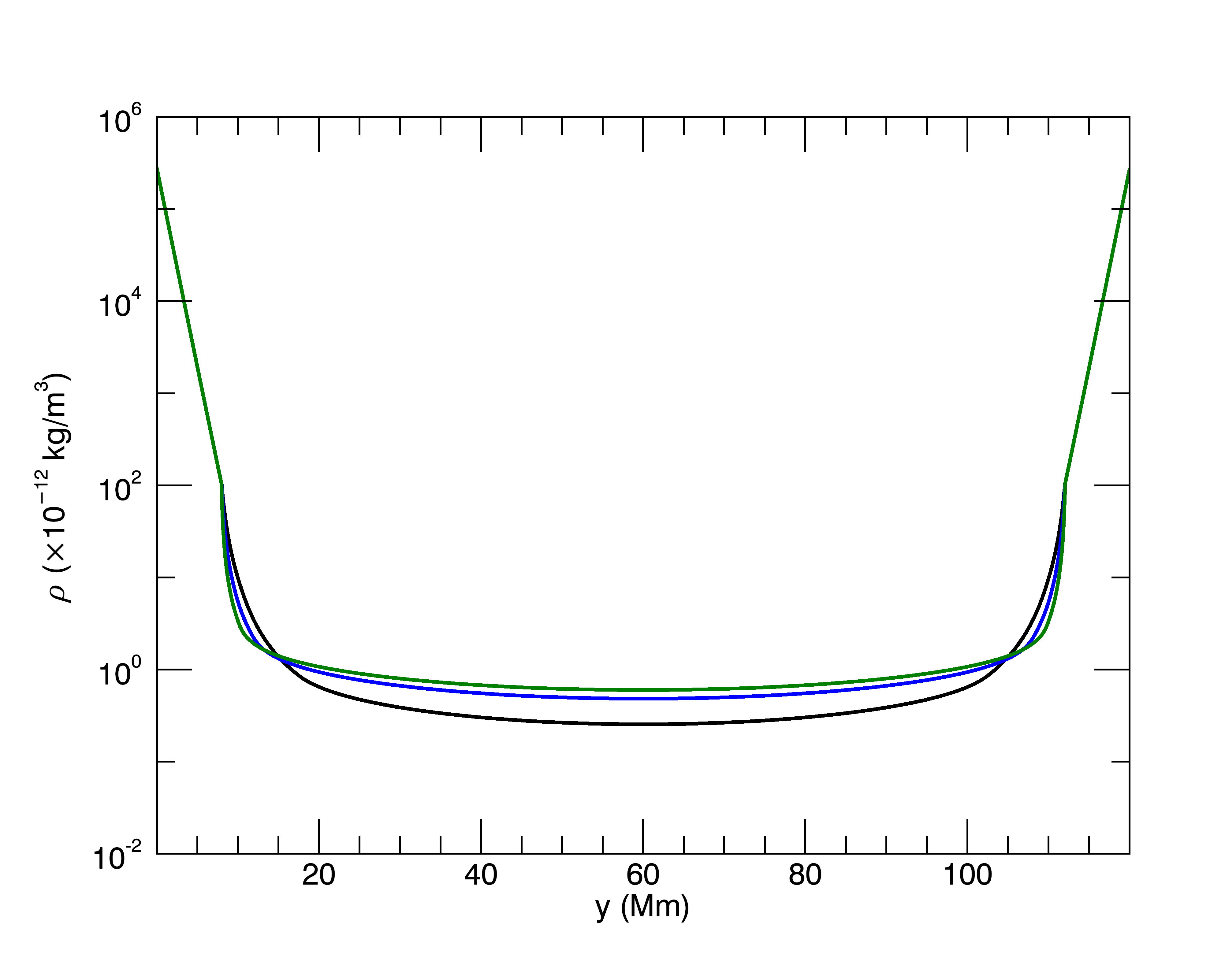}
   \caption{Plot of the field-aligned temperature (left) and density (right) at $x=-2$ Mm (black), $x=-1$ Mm (blue) and $x=0$ Mm (green). The small panel inserted on the temperature plot shows a zoomed version in the TR (between $y=7.5$ Mm and $y = 12$ Mm). The symbols represent the numerical gridpoints.}
   \label{fig_1D_equilibrium}
\end{figure*}

Overplotted (black lines) on these contour plots are the (initial) locations of the chromosphere-TR and TR-corona boundaries. Here, we have defined the top of the TR to be the location where thermal conduction $(\nabla \cdot F_c)$ changes sign (i.e.~changes from an energy loss to an energy gain). The chromospheric boundary is defined as the location where the temperature has decreased to $T = 2 \times 10^4$ K. The locations of these boundaries are determined in the initial setup and are then assumed fixed in all subsequent calculations. For the particular setup studied in this paper, this assumption is reasonable as the modest heating occurring in the corona does not affect the location of the TR and the chromospheric boundaries. 
Figure \ref{fig_1D_equilibrium} shows a (field-aligned) cross-section of the temperature and density at $x=-2$ Mm (exterior), $x=-1$ Mm (shell) and $x=0$ Mm (middle of the loop). It is clear from this figure that the temperature and density are well resolved in the transition region. Indeed, the minimum temperature length scale $L_T$ in the loop is of the order of 125 km, occurring in the lower TR. Thus, the field- aligned grid resolution ($dy \sim 30$ km) is more than adequate to fully resolve the TR. We note here that the temperature at the top of the TRs remains between $0.51-0.83$ MK and, hence, the cutoff temperature $T_c = 0.5$ MK is always located in the TR.

Figure \ref{fig_init_cond2} shows contours of the magnetic field components $B_x$ and $B_y$ as well as the Alfv\'en speed. Before relaxation, the background field is set to be uniform in $y$ with $B_y = 10$ G. Following the (numerical) relaxation, there is a shallow cross-field gradient in $B_y$ as well as slight expansion of the field in the TR (reflected in the appearance of a small $B_x$ component). The Alfv\'en speed shows a similar pattern as the contour of the density in Figure \ref{fig_init_cond1}, varying between $1000-1800$ km/s in the corona, where the highest Alfv\'en speed occurs outside the coronal part of the loop. Finally, we note that our 2D setup results in a plasma beta of the order of 0.01 in the corona.

\begin{figure*}[t]
\centering
\includegraphics[width=0.4\textwidth]{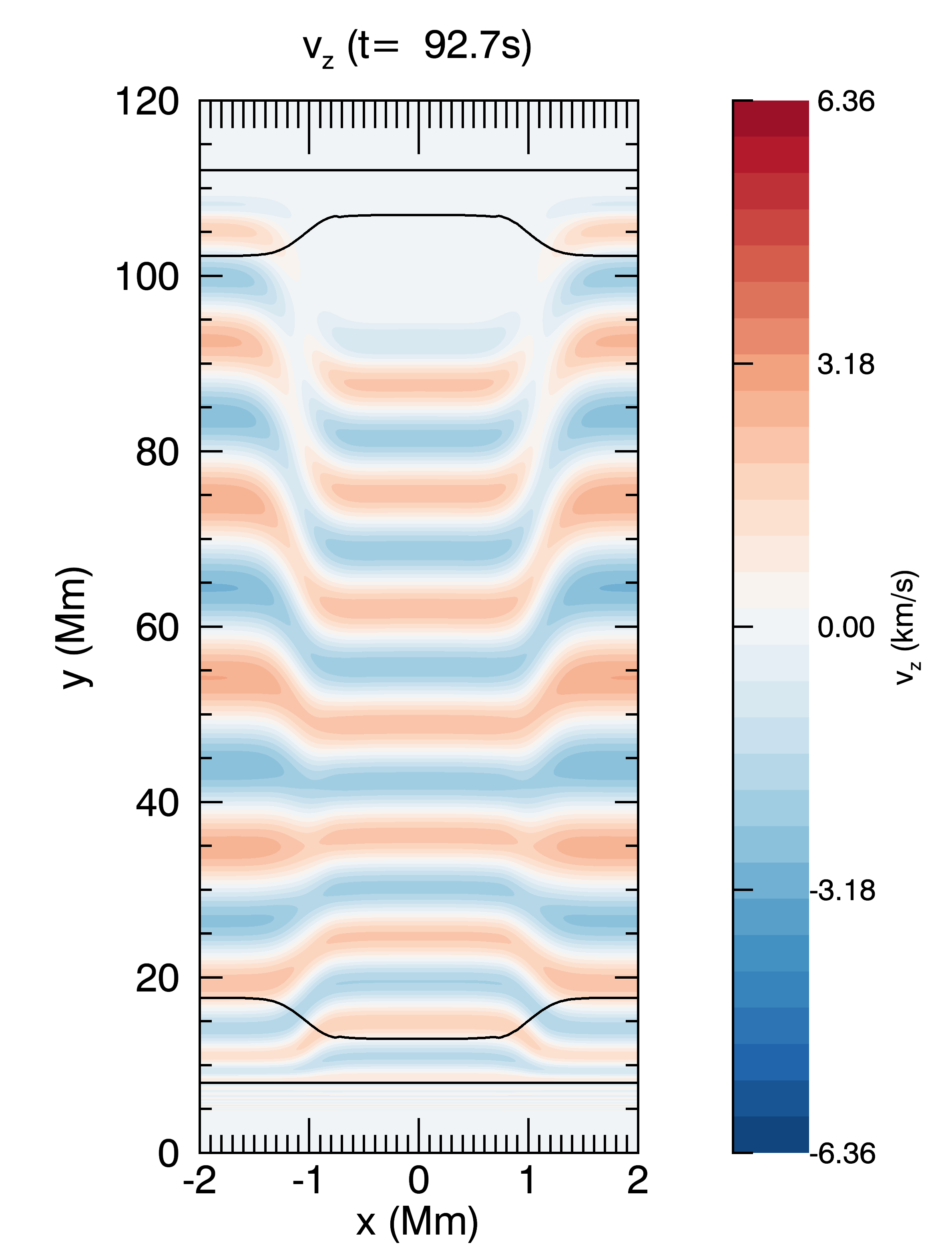}
\includegraphics[width=0.4\textwidth]{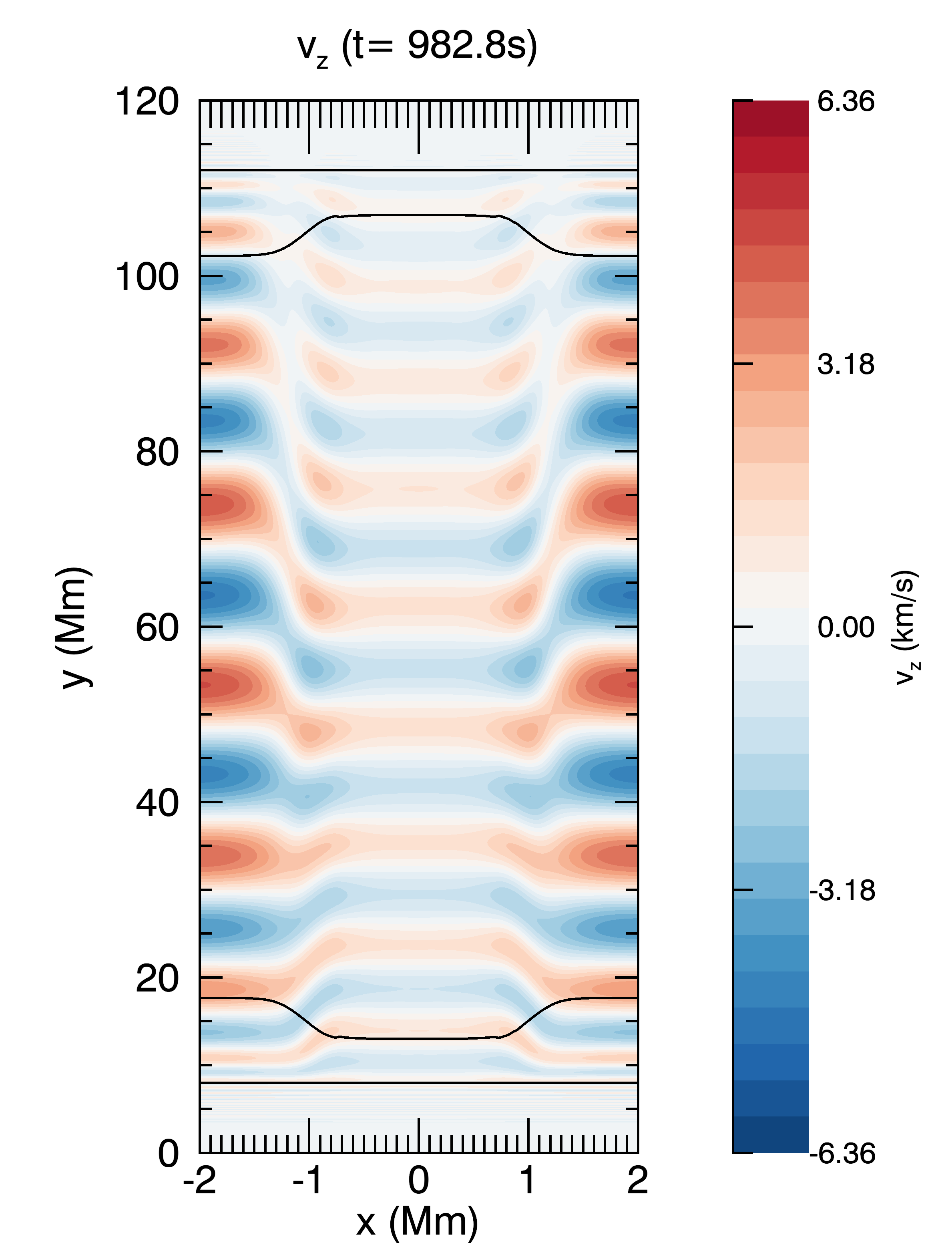}
   \caption{Contour of $v_z$ (km/s) for the viscous simulation at $t=93$s (left panel) and $t=983$s (right panel).}
\label{fig_contour_vz_visc}
\end{figure*}

\subsection{Driver}
We implement a sinusoidal driver at the top of the first chromosphere ($y=7.8$ Mm) in the invariant $z$ direction. The driver is implemented as an external force, $F_z$, in the $z$ component of the momentum equation (Eqn.~\eqref{momentum_eq}) as
\begin{equation}
F_z = -\rho v_0 \omega \cos(\omega t)\,.
 \label{driver}
\end{equation}
Here $v_0 = 0.7$ km/s is 1\% of the local Alfv\'en speed, and $\omega=\frac{2\pi}{P}$ is the angular frequency, where $P \approx 12$s is the period of the driver. This driver generates left and right (up and down) propagating Alfv\'en waves that propagate along the field (in the $y$ direction). Although most observed waves and oscillations in the solar corona have periods of a few minutes (see e.g. \citealt{paper:DeMoortel2012}), in this study, we use a high frequency driver to ensure a sufficient number of wavelengths in the corona to lead to significant phase mixing (see Section \ref{s:Discussion} for further discussion). 

For the parameters considered in our initial conditions, the thermal conduction and optically thin radiation timescales are of the order of 1000-4000 seconds. Therefore, to ensure the effects of thermal conduction and optically thin radiation are captured, we continue to drive the simulation for $t=500P \approx 6000$s.


\section{Results}  
\label{s:Results}

In this paper, we focus on the analysis of a Lare2D simulation  which includes a dynamic viscosity of $\rho\nu=5\times 10^{-4}$ kg m$^{-1}s^{-1}$ but without resistivity (i.e.~$\eta=0$), to avoid diffusion of the background Alfv\'en speed profile. In addition, we will also present results from the corresponding ideal simulation, as well as a simulation which was run for the same total time but without a driver implemented (i.e.~a simulation where the numerical relaxation after imposing the cross-field heating profile continues for the total time interval considered here). 
In Figure \ref{fig_contour_vz_visc} we show contours of $v_z$ for the viscous simulation at $t=93$s (left), when the first pulse generated by the driver reaches the far TR, and $t=983$s (right), representing a later stage of the simulation.
Initially, after a small amount of phase mixing in the lower part of the loop, the Alfv\'en speed profile reverses in the corona and hence this phase mixing is `undone' before more substantial phase mixing occurs close to the far leg of the loop (i.e.~the loop leg furthest away from the driver). Due to the high frequency of the driver, the wavelength of the waves is relatively small in the corona, which allows most of the waves to be transmitted into the transition region. Only a modest amount of wave energy (of the order of 15 \%) is reflected back into the coronal part of the domain. Figure \ref{fig_contour_vz_visc} (right) shows a contour of $v_z$ at $t = 983$s where the oscillation amplitudes are now of the order of 7 km/s (compared to 2-3 km/s at earlier times). The interference of the driven and reflected waves leads to the partial establishment of a standing wave in the coronal part of the domain. When dissipation and phase mixing are absent (the core part of the loop in the ideal simulation), an almost exact standing waves is established after some time. Either in the shell region (where phase mixing takes place) or when viscosity is present, the coronal part of the domain eventually evolves to a steady-state which is a combination of standing and propagating waves. Phase mixing of the driven and reflected waves in the shell region leads to an in intricate pattern of fine scale structuring. In the viscous simulation, this will lead to heating in the shell regions, as can be seen in the left-hand panel of 
Figure \ref{fig_contour_rho_T_visc}, which shows contours of the relative change in temperature, $(T-T_0)/T_0$ at the end of the simulation ($t=5734$s).

\subsection{Long-Period field-aligned flows and oscillations}
\label{sec:upflow}
The aim of this paper is to study any changes in the coronal density profile induced by evaporation of TR or chromospheric plasma, following heating in the corona. However, other field-aligned flows are also present in the domain. In order to be able to distinguish these flows from the evaporation, we now look at a detailed comparison between the ideal and the viscous simulations, as well as the non-driven simulation.

\begin{figure}[!t]
\centering
\includegraphics[width=0.5\textwidth]{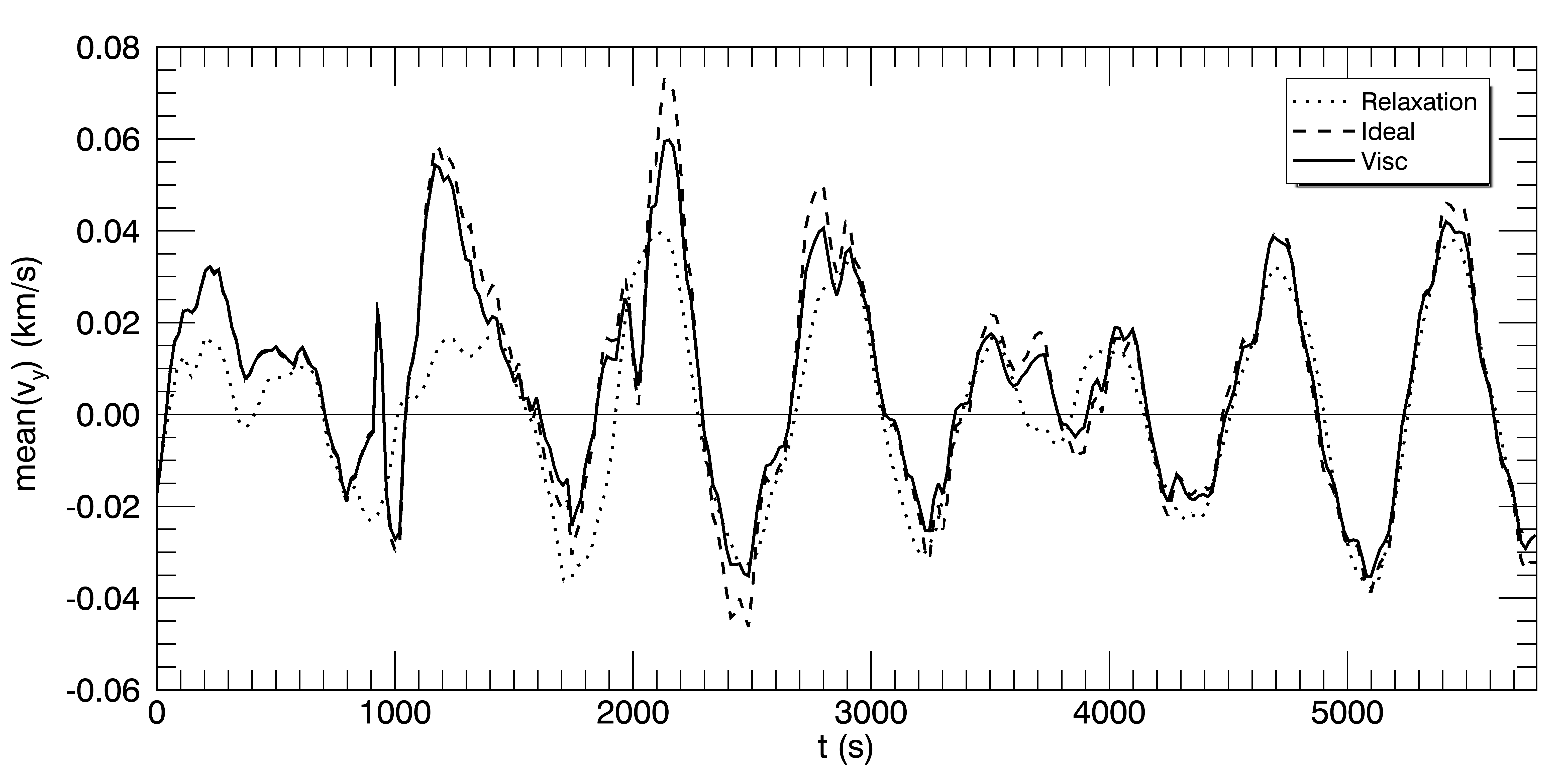}
   \caption{Plot of the mean field-aligned velocity $v_y$ integrated over the first TR-coronal boundary in the central part of the loop ($-1<x<1$ Mm) for the viscous (solid line), ideal (dashed line) and non-driven (dotted line) simulations as a function of time.}
\label{vy_relax_ideal}
\end{figure}

Figure \ref{vy_relax_ideal} shows the field-aligned velocity $v_y$, integrated over the near TR-coronal boundary for the core of the loop ($-1 < x < 1 $ Mm) as a function of time. We can see that a long-period ($P\sim 650$s) oscillation is present in all simulations, which is due to the system still evolving under the imposed cross-field background heating profile. The period is related to the slow travel time in the domain (where the time for a slow wave to travel from the top of the first chromosphere to the top of the far chromosphere in the middle of the domain is about 650s). The oscillations are very small amplitude, of the order of $|v_y|< 0.1$ km/s. Comparing with the non-driven (dotted line) simulation, we can see that the maxima in the ideal simulations (dashed line) are slightly larger. At this TR-coronal boundary, positive values of the field-aligned velocity $v_y$ correspond to an upflow into the corona. These additional upflows are due to the ponderomotive force associated with the Alfv\'en waves generated by the forcing term. A more detailed explanation of these ponderomotive upflows is given using a simple 1D model in Appendix \ref{appendix_pond_force}.  For the viscous simulation (solid line), we find upflows which are slightly smaller than in the ideal case, due to the effect of viscosity on the upflows. Although the ponderomotive upflows are not the main focus of our study, it is worth remarking here that these upflows are likely over-estimated, as we consider the plasma to be fully ionised everywhere in the numerical domain \citep[see e.g.][]{Laming2017}. However, this will equally affect the ponderomotive upflows in all simulations considered here and, hence, does not affect our estimate of the evaporative upflows, which we obtain by comparing the upflows in the ideal and non-ideal simulations.

\subsection{Plasma Changes}
\begin{figure*}[t]
\centering
\includegraphics[width=0.45\textwidth]{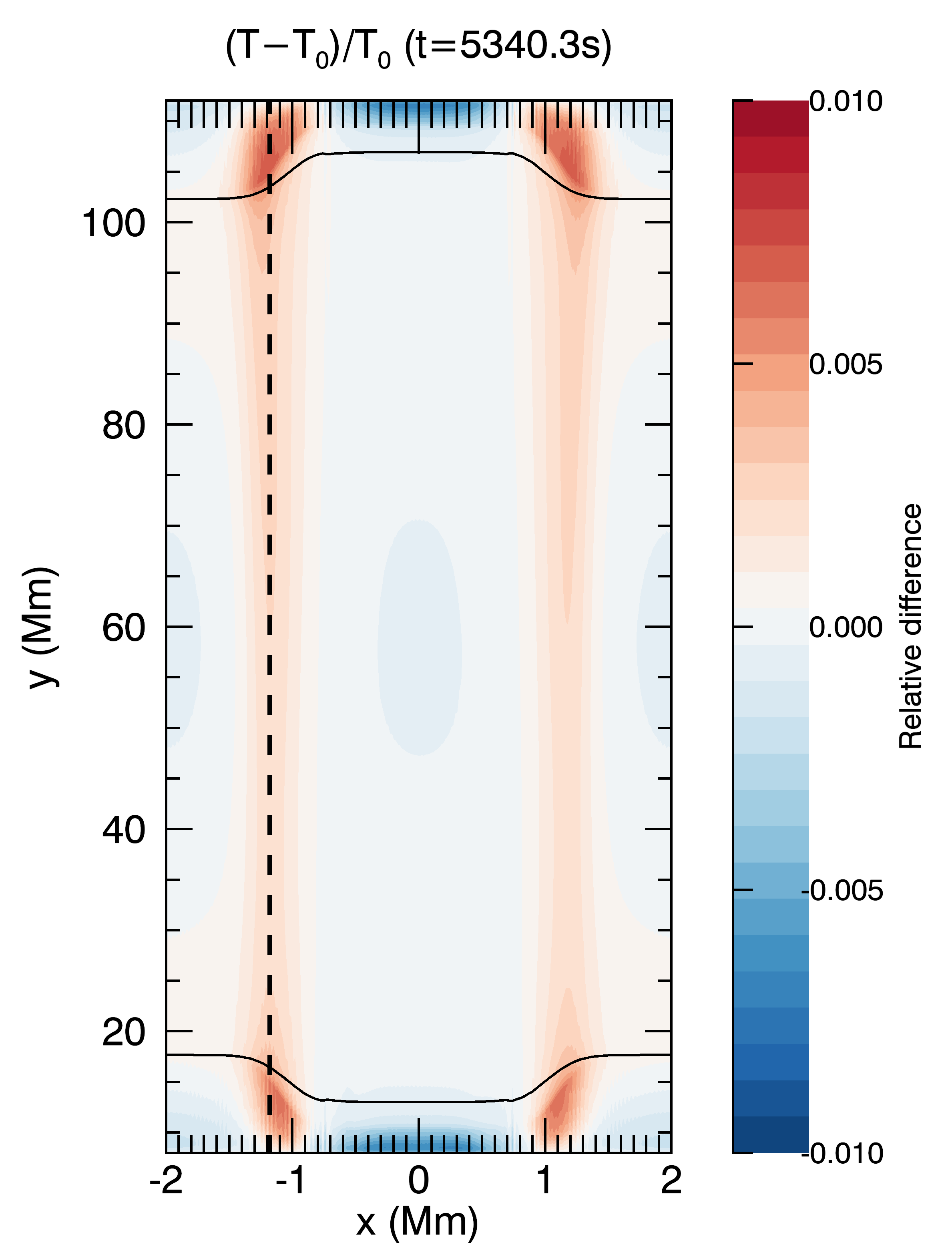}
\includegraphics[width=0.45\textwidth]{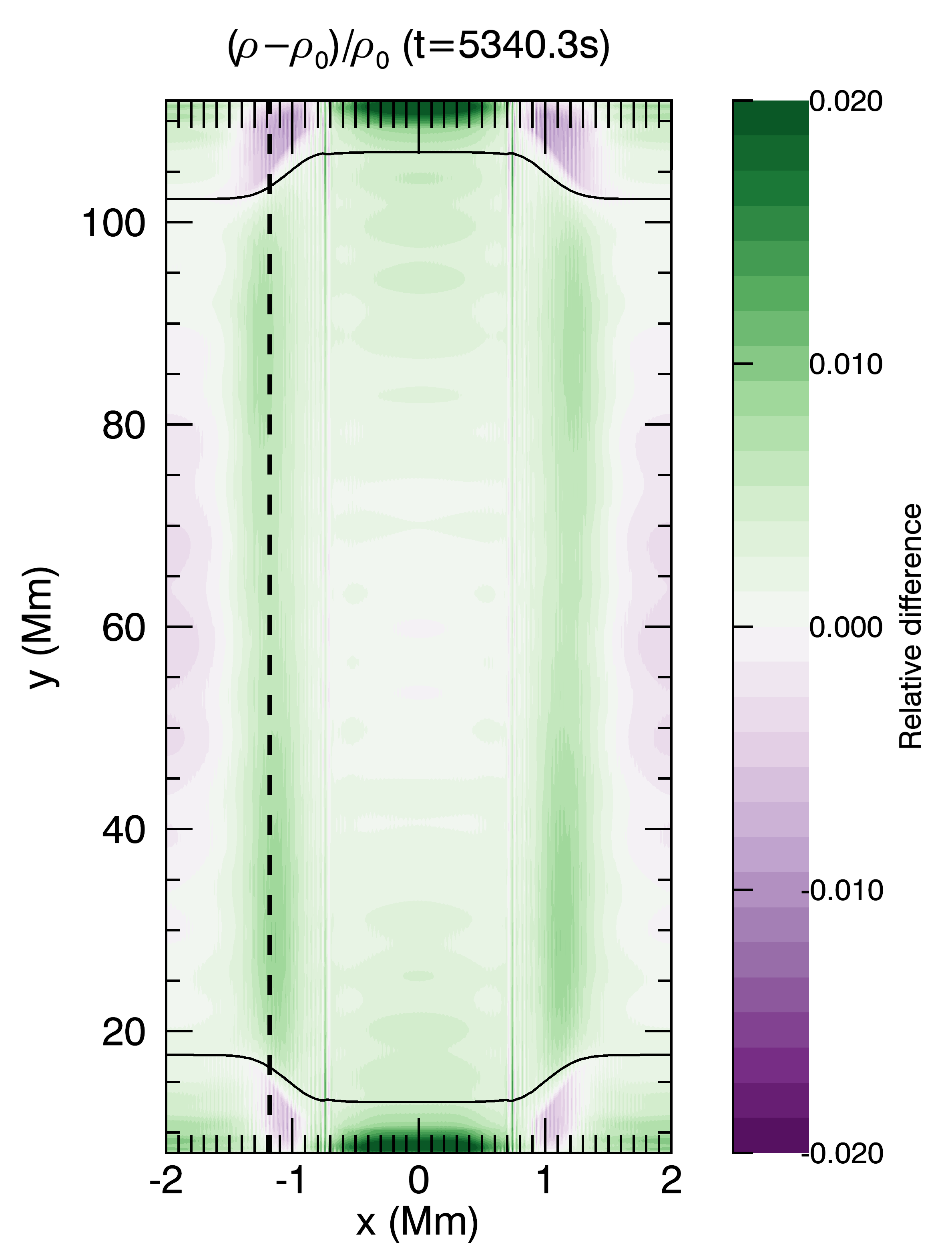}
   \caption{Contours of $(T-T_0)/T_0$ (left) and $(\rho - \rho_0)/\rho_0$ (right) at $t=5340$s for the viscous simulation.}
\label{fig_contour_rho_T_visc}
\end{figure*}

As already mentioned earlier, phase mixing in the shell regions leads to dissipation of the wave energy in the viscous simulations (Figure \ref{fig_contour_vz_visc}) and hence to an increase in the local temperature (left-hand panel of Figure \ref{fig_contour_rho_T_visc}). At early times, the dissipation occurs in the far coronal part of the loop ($z > 60$ Mm) as can be expected from the phase mixing pattern in $v_z$ (LHS of Figure \ref{fig_contour_vz_visc}). At later times, when a steady state is established, a small increase in the relative temperature all along the shell regions can be observed, although the increase remains slightly higher in the far leg of the loop at all times. The actual increase in the coronal temperature is very small (of the order of 4000K or less than 0.5\% of the background coronal temperature) due to the limited energy input, which results from the combination of the relatively small amplitude of the wave driver and the limited reflection at the TR-coronal boundary. In other words, any wave energy that has not yet been dissipated by the time the propagating waves reach this boundary is mostly transmitted to the lower atmosphere rather than being contained in the coronal part of the loop. There is no noticeable change in temperature in the core (coronal) part of the loop and the loop environment where phase mixing does not occur. 

\begin{figure*}[t]
\centering
\includegraphics[width=0.85\textwidth]{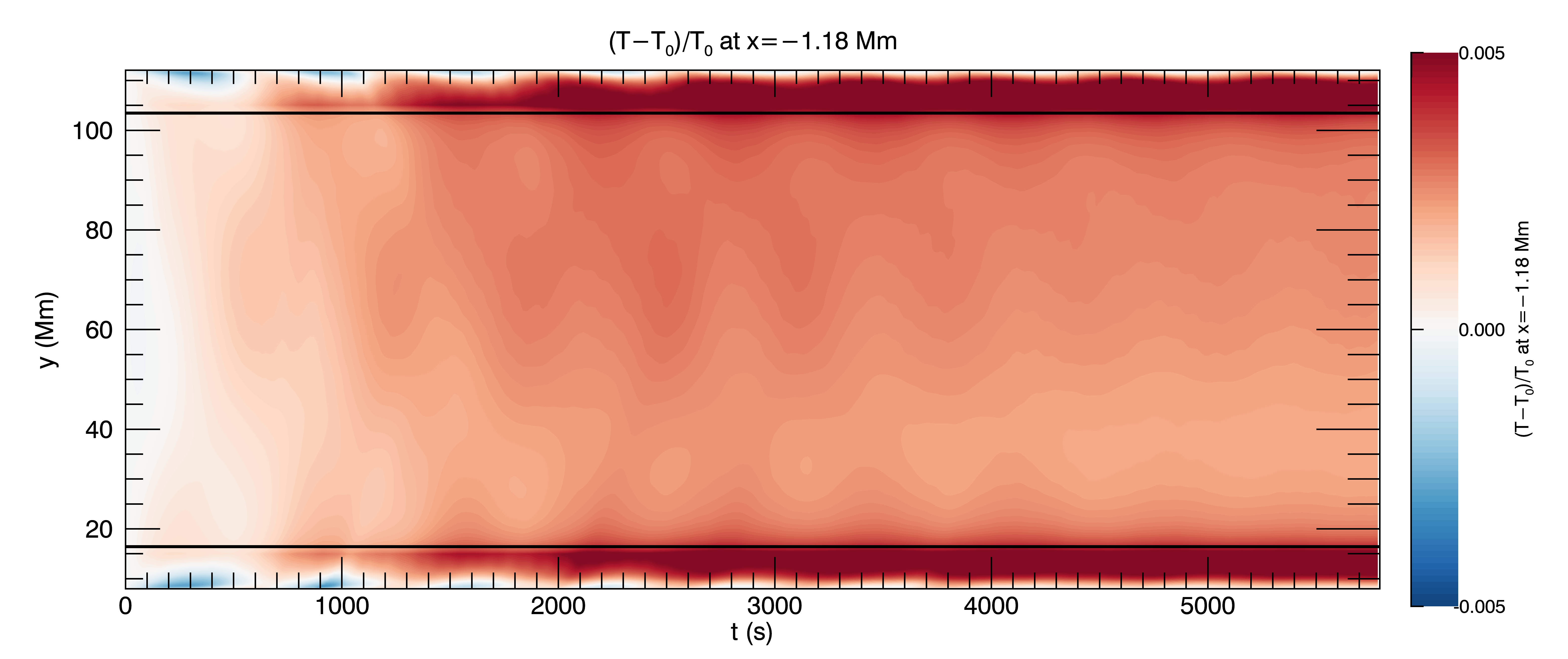}
\includegraphics[width=0.85\textwidth]{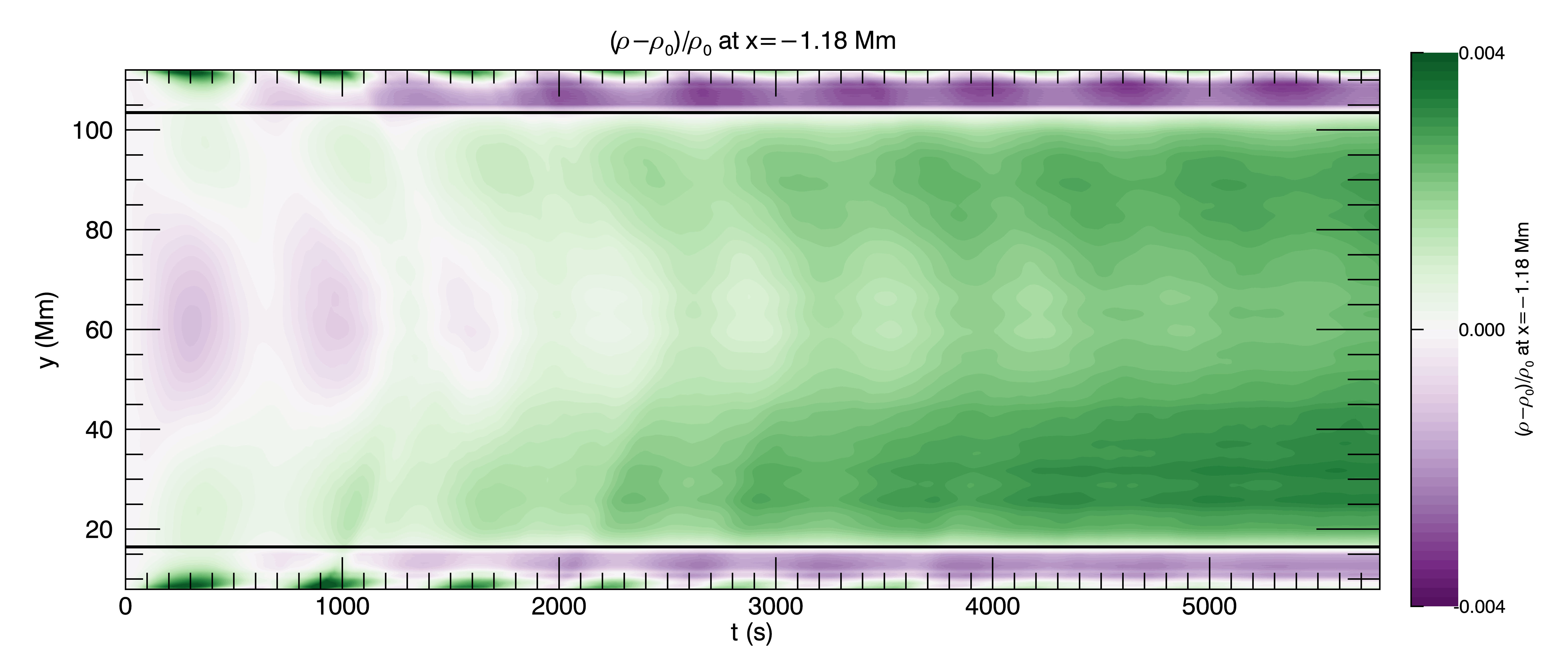}
  \caption{Time distance graphs of the relative changes in temperature (top) and density (bottom) along x = -1.18 Mm (marked by vertical dashed line in Figure \ref{fig_contour_rho_T_visc}).}
\label{time_distance_T_rho}
\end{figure*}

\begin{figure*}[t]
\centering
\includegraphics[width=0.45\textwidth]{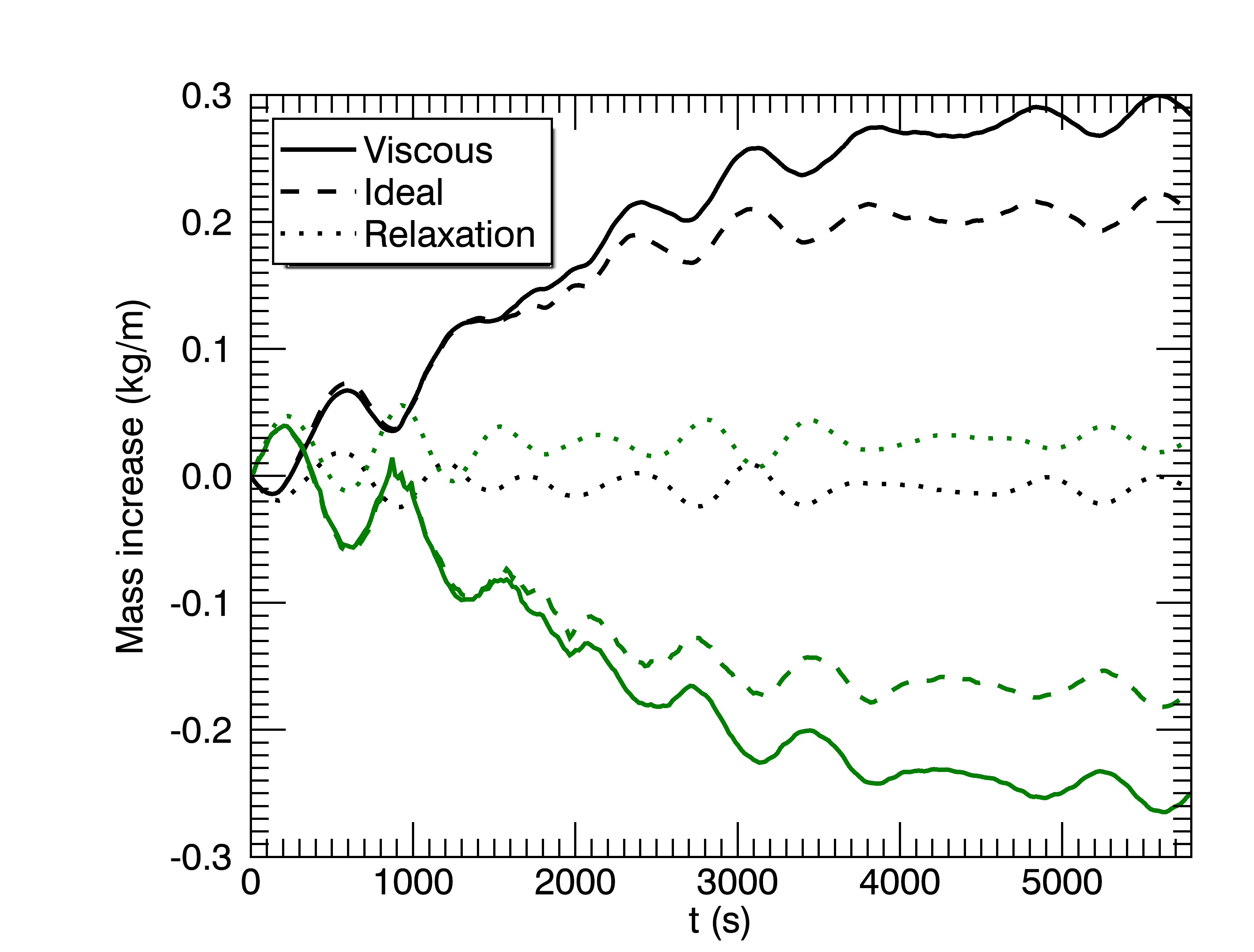}
\includegraphics[width=0.45\textwidth]{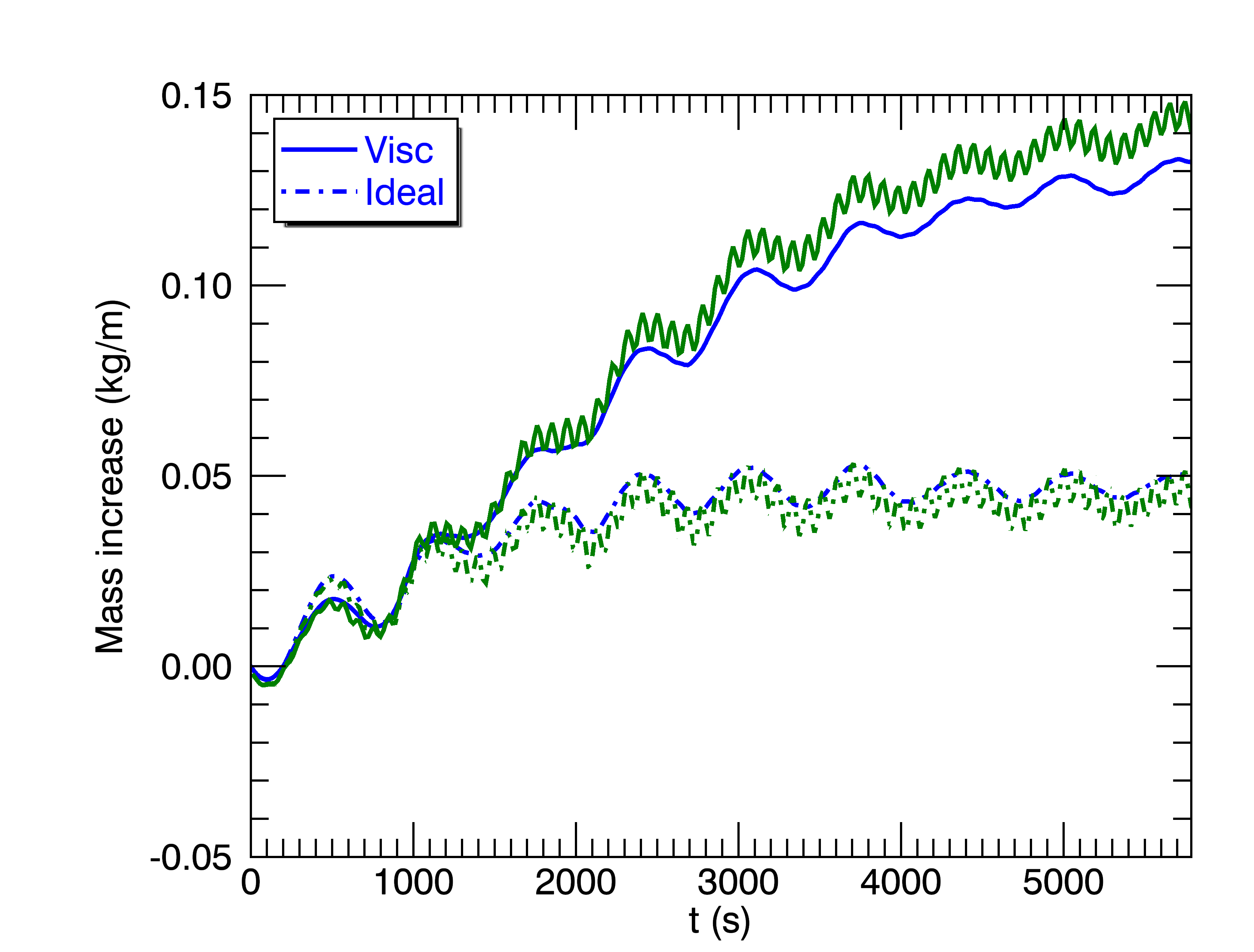}
   \caption{(Left) Plot of the mass increase (kg/m) for the three simulations (viscous, solid lines; ideal, dashed lines; non-driven, dotted lines). The black lines correspond to the integrated mass in the coronal part of the domain whereas the green lines represent the mass in the non-coronal part. (Right) The mass increase (kg/m) (blue lines) but now only integrated over the coronal part of the shell region (viscous, solid lines; ideal, dotted-dashed lines). The green lines represent the time integrated, averaged mass flux (where the averaging is done over the 4 boundaries of the shells).}
\label{intmass_relax_ideal_visc}
\end{figure*}

The right-hand panel of Figure \ref{fig_contour_rho_T_visc} shows the relative change in the density at the end of the simulation. Again, the most significant changes in the coronal part of the loop are limited to the shell region, where an increase of about 1\% in the density has occurred (with a corresponding decrease in the TR and chromosphere of the shell region). However, as described above in Section \ref{sec:upflow}, evaporation due to (coronal) heating is only part of this change in density, as upflows due to the ponderomotive force associated with the Alfv\'en wave driver and the ongoing evolution of the background due to the enhanced heating profile are also present. Given that the upflows due to the background evolution are oscillatory (dashed line in Figure \ref{vy_relax_ideal}), it is likely that this process does not contribute significantly to the relative density change in the corona. This is indeed confirmed in Figure \ref{intmass_relax_ideal_visc} where we see that there is essentially no change in mass in the corona for the non-driven simulation (dotted black line).

In Figure \ref{time_distance_T_rho}, we show time-distance diagrams of the change in relative temperature (top) and density (bottom) along a cross section in the middle of the shell region ($x=-1.18$ Mm, marked by the vertical dashed lines in Figure \ref{fig_contour_rho_T_visc}). A few features immediately stand out in the evolution of the relative temperature changes. First of all, the time-distance plot of the relative temperature confirms that the heating is indeed  asymmetric, with the relative increase in temperature slightly larger in the far leg of the loop ($y>60$ Mm). Because of the specific Alfv\'en speed profile, phase mixing only develops significantly in the loop leg further away from the driver and hence the wave dissipation is more efficient in the far leg. Secondly, there are clear periodic changes in the temperature, propagating upwards from the TR into the corona with a speed of about 120 km/s (estimated from the slope of the diagonals in the contour plot). A comparison with Figure \ref{vy_relax_ideal} shows that this periodicity must be associated with the ongoing evolution induced by the heating profile, as the same periods are also present in the field-aligned flows in both the non-driven and ideal simulations. As the perturbations associated with both this ongoing relaxation and the ponderomotive force are adiabatic, no net heating (i.e.~integrated over time) is expected.  Indeed, this is confirmed by analy4 the corresponding time-distance plot for the ideal simulation (not shown), where periodic changes in the temperature are present in the TRs but there is no net increase of temperature over time. For the viscous simulation on the other hand, there is a clear net increase in the temperature over time in both the coronal part of the domain and the lower atmosphere.

Focusing on the first 1000 seconds of the relative temperature change in the viscous simulation, there is a first small heating event occurring roughly between $y=60-80$ Mm at $t=500-800$ s. Comparing with the contour plots of $v_z$ (the driven Alfv\'en waves) in Figure \ref{fig_contour_vz_visc}, it is clear that this heating is a direct result of the viscous dissipation of the phase-mixed Alfv\'en waves. Following this heating event, a thermal conduction front can be seen in the form of a downward propagating increase in temperature (i.e.~towards higher values of $y$) with a steep gradient (high velocity). There is some evidence of the temperature change also extending towards the other loop leg ($y < 60$ Mm) although this is harder to see. Similar heating `events' then appear to re-occur at intervals of about 600-700 s. However, this periodicity is essentially a modulation of the heating from phase mixing by the long-period oscillations discussed above. Although it is not easy to see the downward propagating change in temperature due to conduction, some instances can be identified where the contours are tilted towards the right in the upper half of the domain, which corresponds to propagation from the corona to the TR in the far leg of the loop ($y > 60$ Mm). Finally, given the background Alfv\'en speed profile, phase mixing of the driven Alfv\'en waves will also occur in the TRs and, hence, direct heating in these regions is expected. Therefore the overall, net increase in temperature in the TRs over time is a combination of direct heating and the downward conduction of coronal changes in the temperature. 

The bottom panel in Figure \ref{time_distance_T_rho} shows the corresponding time-distance diagram for the relative change in density. There are clear upflows present propagating from the lower parts of the loop up into the corona. Again, the periodicity of these density changes is associated with the background long-period oscillations as discussed in Section \ref{sec:upflow}. The upflows are caused by two effects: the ponderomotive force associated with the driven Alfv\'en waves and the evaporation from the lower atmosphere into the corona following (coronal) heating by phase mixing of the Alfv\'en waves. Indeed, comparing with the ideal simulation (where only the ponderomotive mass increase is present) does reveal that for the viscous simulation, there is a larger increase in mass in the shell regions, which is associated with the evaporation from the lower atmosphere. 

To confirm this, we look at the change in mass integrated over time in Figure \ref{intmass_relax_ideal_visc}. The right-hand panel shows the change in mass integrated over the coronal part of the shell regions only, for the viscous (solid blue line) and the ideal (dashed blue line) simulation. An increase in mass is indeed present in both cases but the increase is very clearly larger in the viscous simulation. This excess in mass increase is associated with the evaporative upflows from the lower atmosphere induced by the Alfv\'en wave heating. The green lines correspond to the time integrated, averaged mass flux (where the averaging is done over the 4 TR-coronal boundaries of the shell regions) and, hence, confirm that the mass increase in the coronal part of the shell is indeed due to mass flux inflowing from the lower atmosphere.

The left-hand panel of Figure \ref{intmass_relax_ideal_visc} shows the mass changes for the full loop (i.e.~including the loop core and the environment outside the loop) for the three simulations with the coronal mass change plotted in black and the lower atmosphere in green. This figure essentially summarises the findings described above. Looking at the non-driven simulation (dotted line) confirms there is no significant change in mass in the corona or lower part of the domain due to the oscillatory field-aligned perturbations associated with the continued (background) evolution. For the ideal and viscous simulations, there is a clear increase of mass in the coronal part of the loop (with a corresponding decrease in the lower atmosphere). For the full domain (rather than just the shell region as in the right-hand panel), the mass increase is dominated by the ponderomotive effect, given the relatively small difference between the mass increase for the viscous (solid line) and ideal (dashed line) simulations. Saying that, the viscosity also acts on the ponderomotive upflows (see Figure \ref{vy_relax_ideal}) and, hence, the effect of the evaporative upflows is somewhat larger than indicated by the direct difference between the two simulations.

We can estimate the magnitude of the evaporative upflows by considering the difference in the mean upflows through the TR-coronal boundaries of the shell regions between the viscous and ideal simulations. We find that this evaporative upflow is of the order of $5-20$ m/s and is approximately equal on both the near (closest to the driver) and far TR-coronal boundaries. Using the continuity equation and the observed mass increase in the shell regions confirms that this evaporative upflow does indeed match the required average evaporation.

\begin{figure}[t]
\centering
 \includegraphics[width=0.45\textwidth]{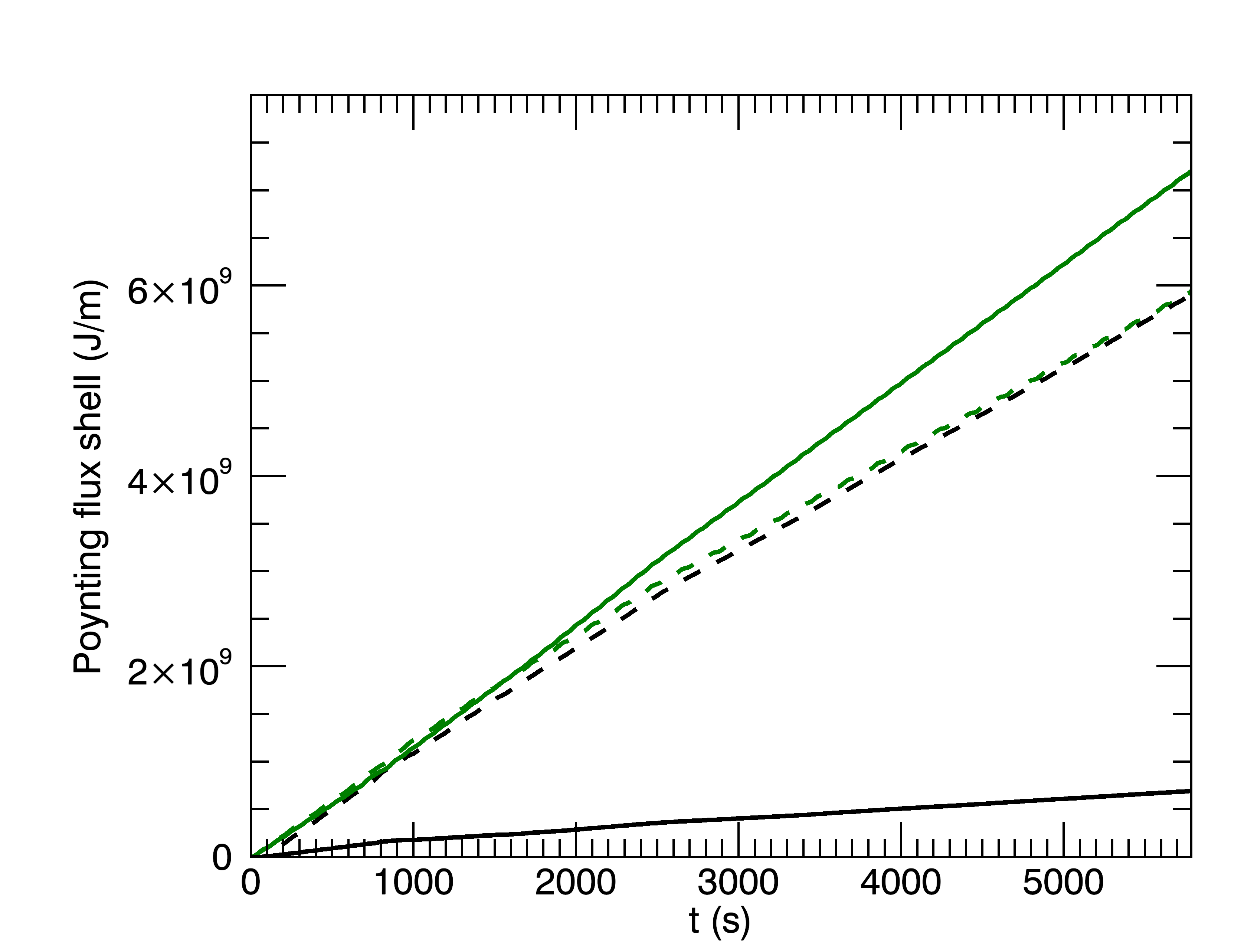} 
\caption{Plot of the time integrated Poynting flux $\int_0^t \int\left(E\times B\right)_y dSdt$ (J/m), with $S$ the boundaries of the left shell, for the viscous (solid lines) and ideal simulation (dashed lines). The green lines represent the lower TR-coronal boundary and the black lines the upper boundary.}
\label{fig_poynting_flux}
\end{figure}

Figure \ref{fig_poynting_flux} shows the time and volume integrated vertical component of the Poynting flux $\int_0^t \int\left(E\times B\right)_y dSdt$ (J/m), over the TR-coronal boundaries of one of the shell regions of the loop. The green lines represent the near (closest to the driver) TR-coronal boundary and the black lines the far boundary. Firstly, as expected for the ideal simulation, the Poynting flux entering the coronal shell region (through the near boundary - green dashed line) is the same as the Poynting flux leaving this region (black dashed line). For the viscous simulation on the other hand, we can see that the Poynting flux leaving the coronal shell region (solid black line) is substantially smaller than the Poynting flux entering the region (green dashed line). This difference is of the same order as the internal energy increase in the shell region (not shown here). The Poynting flux entering the shell region through the near TR-coronal boundary (green lines) differs in the ideal and viscous simulation due to the different mixture of propagating and standing waves in this region. For the ideal simulation, the standing wave component dominates, which reduces the amount of Poynting flux through the near boundary, as the standing regime creates a node in $v_z$ \citep[see e.g.][]{Prokopyszyn2019}. Comparing with coronal energy requirements, the difference between the average Poynting flux entering and leaving through the TR-coronal boundaries of the shell region for the viscous simulation is of the order of $1.5-2$ Jm$^{-2}$s$^{-1}$, several orders of magnitude too low, even for the Quiet Sun ($\sim 3\times 10^2$ Jm$^{-2}$s$^{-1}$, \citealt{paper:Withbroe1977}). 

\section{Discussion and Conclusions} \label{s:Discussion}

In this paper, we use 2.5D numerical simulations to investigate whether a pre-existing density profile is modified by the evaporative upflows following (viscous) heating by phase mixing of Alfv\'en waves. The Alfv\'en waves are driven through an additional forcing term in the momentum equation. The simulations account for the effects of gravitational stratification, thermal conduction, and optically thin radiation. A fully resolved atmospheric model (including the chromosphere as a mass reservoir) is achieved with relatively modest resolution by artificially broadening the TR using the approach proposed by \cite{paper:Lionello2009} and \cite{paper:Mikic2013}. To include a transverse density profile (perpendicular to the background magnetic field) in the initial setup, a heating function which varies in the cross-field direction is imposed. 

Throughout the simulations (running for a total time of more than 5000 seconds), a complex combination of Alfv\'en waves ($v_z$) and longitudinal perturbations ($v_y$) is present in the domain. The longitudinal perturbations are a combination of long-period oscillations resulting from the ongoing evolution due to the imposed cross-field heating profile, the ponderomotive effects associated with the driven Alfv\'en waves, and the evaporative upflows resulting from heating occurring in the coronal part of the shell regions of the loop due to phase mixing of the Alfv\'en waves. 

By comparing with the results from the ideal simulations, we are able to distinguish the evaporative upflows present in the viscous simulations from the other (ideal) longitudinal perturbations. This allows us to identify the change in mass in the shell regions of the coronal loop which results from the viscous heating by phase mixing of Alfv\'en waves. For the particular setup studied in this paper, the amount of heating through viscous dissipation of the phase-mixed Alfv\'en waves in the corona is extremely small and, hence, the evaporative upflows associated with this heating are insignificant. Therefore, in this case, the heating-evaporation cycle has no noticeable effect on the transverse density profile (or Alfv\'en speed profile) which causes the Alfv\'en wave phase mixing. 

As remarked above, the choice of our high-frequency driver ($P \sim 12$ s) implies that a substantial amount of the wave energy is transmitted down to the far TR and chromosphere. Indeed, the short wavelengths associated with the high-frequency driver allow the rapid development of phase mixing in the shell regions of the loop but, at the same time, a substantial amount of energy (about 20-25\%) is not dissipated by the time the waves reach the far TR and is lost to the lower atmosphere \citep[see e.g.][]{Hollweg1984b, Hollweg1984a, Berghmans1995, DePontieu2001}. This implies that simply increasing the amplitude of the driven Alfv\'en wave would only have a limited effect, as a substantial amount of the wave energy would still be lost from the coronal volume, even in the shell regions. In the core of the loop, almost all energy (of the order of 85\%) is transmitted to the far TR and chromosphere. Although the TR has been broadened somewhat to ensure sufficient numerical resolution (see Section \ref{s:Model}), this broadening does not substantially affect the amount of wave energy that is transmitted and reflected. We also note here that the method proposed by \cite{paper:Lionello2009} and \cite{paper:Mikic2013} conserves the total amount of energy that is delivered to the chromosphere and that although there can be small differences with the flows in the modified region (i.e.~where $T < T_c$), the mass flux into the corona is preserved \citep{Johnston2020}. Therefore, the artificial broadening does not have a significant effect on the evaporative upflows in this study.

In a forthcoming paper, we will extend the study presented here to the low frequency limit, as waves and oscillations observed so far in the solar corona mostly have periods on the order of a few minutes \citep[e.g.][]{paper:DeMoortel2012}. This will lead to longer wavelengths and, hence, the waves would experience more reflection at the boundaries between the coronal part and the lower atmosphere which would, in turn, allow for more energy to be contained in the coronal part of the entire loop. In the initial phase of propagation, there would be less pronounced phase mixing for longer wavelength waves but, over time, phase mixing (in time) would lead to large gradients and, hence, dissipation of the wave energy in the coronal part of the shell regions of the loop. However, longer wavelength waves would take longer to phase mix and therefore, even if more energy is retained in the corona, it remains to be investigated whether this can result in more energy converted into heating and, if so, whether this results in stronger evaporative upflows on relevant timescales.

Even in the long wavelength limit, the simulations would still be highly idealised and mostly a theoretical study of the evaporation. Several other aspects could be included to make the model more representative of actual coronal loops. For example, in the current study we have only considered viscous heating, and neglected the effect of resistivity. As the resistivity is expected to be very low in the solar corona, this might be acceptable, but it would still be useful to examine the effect of including resistivity. In particular, it would be instructive to study how including resistivity balances the effects of changing the background Alfv\'en speed profile (through diffusing the background magnetic field configuration) and the stronger heating resulting from the additional resistive heating. Further possibilities include a more realistic broadband driver to establish how this affects the energy input into the corona as well as a magnetic field configuration with concentrated sources, where stronger divergence of the magnetic field could enhance the phase mixing process \citep[e.g.][]{paper:DeMoortel2000}.

Finally, the study presented here mostly maintains the background initial conditions (i.e.~the actual loop profile) through the presence of an imposed artificial background heating function. Without the presence of the background heating, \cite{paper:Cargill2016} argued that the thermal evolution (i.e.~the loop cooling) would lead to significant changes in the cross-field density profile (mostly due to draining) on timescales quicker than the heating provided by phase mixing of Alfv\'en waves. Although heating by phase mixing of Alfv\'en waves does not appear efficient for the parameters used in this current study, it does not provide a definitive answer to the question whether phase mixing of Alfv\'en waves can be a self-consistent, stand-alone heating mechanism of solar coronal loops due to the limitations of our model setup described above.

\begin{acknowledgements}
This work has received support from the UK Science and Technology Facilities Council (Consolidated Grant ST/K000950/1), the European Union Horizon 2020 research and innovation programme (grant agreement No. 647214) and the Research Council of Norway through its Centres of Excellence scheme, project number 262622.
\end{acknowledgements}

\bibliographystyle{aa}        
\bibliography{CE_PM.bib}           

\begin{thebibliography}{52}
\expandafter\ifx\csname natexlab\endcsname\relax\def\natexlab#1{#1}\fi

\bibitem[{{Arber} {et~al.}(2001){Arber}, {Longbottom}, {Gerrard}, \&
  {Milne}}]{paper:Arber2001}
{Arber}, T.~D., {Longbottom}, A.~W., {Gerrard}, C.~L., \& {Milne}, A.~M. 2001,
  Journal of Computational Physics, 171, 151

\bibitem[{{Arregui}(2015)}]{paper:Arregui2015}
{Arregui}, I. 2015, Philosophical Transactions of the Royal Society of London
  Series A, 373, 20140261

\bibitem[{{Arregui} {et~al.}(2012){Arregui}, {Oliver}, \&
  {Ballester}}]{Arregui2012}
{Arregui}, I., {Oliver}, R., \& {Ballester}, J.~L. 2012, Living Reviews in
  Solar Physics, 9, 2

\bibitem[{{Banerjee} {et~al.}(2007){Banerjee}, {Erd{\'e}lyi}, {Oliver}, \&
  {O'Shea}}]{Banerjee2007}
{Banerjee}, D., {Erd{\'e}lyi}, R., {Oliver}, R., \& {O'Shea}, E. 2007,
  \solphys, 246, 3

\bibitem[{{Berghmans} \& {de Bruyne}(1995)}]{Berghmans1995}
{Berghmans}, D. \& {de Bruyne}, P. 1995, \apj, 453, 495

\bibitem[{{Bradshaw} \& {Cargill}(2013)}]{paper:Bradshaw2013}
{Bradshaw}, S.~J. \& {Cargill}, P.~J. 2013, \apj, 770, 12

\bibitem[{{Brooks}(2019)}]{Brooks2019}
{Brooks}, D.~H. 2019, \apj, 873, 26

\bibitem[{{Brooks} {et~al.}(2012){Brooks}, {Warren}, \&
  {Ugarte-Urra}}]{Brooks2012}
{Brooks}, D.~H., {Warren}, H.~P., \& {Ugarte-Urra}, I. 2012, \apjl, 755, L33

\bibitem[{{Brooks} {et~al.}(2009){Brooks}, {Warren}, {Williams}, \&
  {Watanabe}}]{Brooks2009}
{Brooks}, D.~H., {Warren}, H.~P., {Williams}, D.~R., \& {Watanabe}, T. 2009,
  \apj, 705, 1522

\bibitem[{{Cargill}(1994)}]{paper:Cargill1994}
{Cargill}, P.~J. 1994, \apj, 422, 381

\bibitem[{{Cargill} {et~al.}(2016){Cargill}, {De Moortel}, \&
  {Kiddie}}]{paper:Cargill2016}
{Cargill}, P.~J., {De Moortel}, I., \& {Kiddie}, G. 2016, \apj, 823, 31

\bibitem[{{Dadashi} {et~al.}(2011){Dadashi}, {Teriaca}, \&
  {Solanki}}]{Dadashi2011}
{Dadashi}, N., {Teriaca}, L., \& {Solanki}, S.~K. 2011, \aap, 534, A90

\bibitem[{{Dadashi} {et~al.}(2012){Dadashi}, {Teriaca}, {Tripathi}, {Solanki},
  \& {Wiegelmann}}]{Dadashi2012}
{Dadashi}, N., {Teriaca}, L., {Tripathi}, D., {Solanki}, S.~K., \&
  {Wiegelmann}, T. 2012, \aap, 548, A115

\bibitem[{{De Moortel} {et~al.}(2000){De Moortel}, {Hood}, \&
  {Arber}}]{paper:DeMoortel2000}
{De Moortel}, I., {Hood}, A.~W., \& {Arber}, T.~D. 2000, \aap, 354, 334

\bibitem[{{De Moortel} \& {Nakariakov}(2012)}]{paper:DeMoortel2012}
{De Moortel}, I. \& {Nakariakov}, V.~M. 2012, Philosophical Transactions of the
  Royal Society of London Series A, 370, 3193

\bibitem[{{De Pontieu} {et~al.}(2001){De Pontieu}, {Martens}, \&
  {Hudson}}]{DePontieu2001}
{De Pontieu}, B., {Martens}, P.~C.~H., \& {Hudson}, H.~S. 2001, \apj, 558, 859

\bibitem[{{Del Zanna}(2008)}]{DelZanna2008}
{Del Zanna}, G. 2008, \aap, 481, L49

\bibitem[{{Feldman} {et~al.}(2011){Feldman}, {Dammasch}, \&
  {Doschek}}]{Feldman2011}
{Feldman}, U., {Dammasch}, I.~E., \& {Doschek}, G.~A. 2011, \apj, 743, 165

\bibitem[{{Guerreiro} {et~al.}(2013){Guerreiro}, {Hansteen}, \& {De
  Pontieu}}]{Guerreiro2013}
{Guerreiro}, N., {Hansteen}, V., \& {De Pontieu}, B. 2013, \apj, 769, 47

\bibitem[{{Hansteen} {et~al.}(2010){Hansteen}, {Hara}, {De Pontieu}, \&
  {Carlsson}}]{Hansteen2010}
{Hansteen}, V.~H., {Hara}, H., {De Pontieu}, B., \& {Carlsson}, M. 2010, \apj,
  718, 1070

\bibitem[{{Heyvaerts} \& {Priest}(1983)}]{paper:Priest1983}
{Heyvaerts}, J. \& {Priest}, E.~R. 1983, \aap, 117, 220

\bibitem[{{Hollweg}(1984{\natexlab{a}})}]{Hollweg1984b}
{Hollweg}, J.~V. 1984{\natexlab{a}}, \solphys, 91, 269

\bibitem[{{Hollweg}(1984{\natexlab{b}})}]{Hollweg1984a}
{Hollweg}, J.~V. 1984{\natexlab{b}}, \apj, 277, 392

\bibitem[{{Jess} {et~al.}(2015){Jess}, {Morton}, {Verth}, {Fedun}, {Grant}, \&
  {Giagkiozis}}]{Jess2015}
{Jess}, D.~B., {Morton}, R.~J., {Verth}, G., {et~al.} 2015, \ssr, 190, 103

\bibitem[{{Johnston} \& {Bradshaw}(2019)}]{JohnstonBrad2019}
{Johnston}, C.~D. \& {Bradshaw}, S.~J. 2019, \apjl, 873, L22

\bibitem[{{Johnston} {et~al.}(2019){Johnston}, {Cargill}, {Antolin}, {Hood},
  {De Moortel}, \& {Bradshaw}}]{Johnston2019}
{Johnston}, C.~D., {Cargill}, P.~J., {Antolin}, P., {et~al.} 2019, \aap, 625,
  A149

\bibitem[{{Johnston} {et~al.}(2020){Johnston}, {Cargill}, {Hood}, {De Moortel},
  {Bradshaw}, \& {Vaseekar}}]{Johnston2020}
{Johnston}, C.~D., {Cargill}, P.~J., {Hood}, A.~W., {et~al.} 2020, arXiv
  e-prints, arXiv:2002.01887

\bibitem[{{Johnston} {et~al.}(2017a){Johnston}, {Hood}, {Cargill}, \& {De
  Moortel}}]{paper:Johnston2017a}
{Johnston}, C.~D., {Hood}, A.~W., {Cargill}, P.~J., \& {De Moortel}, I. 2017a,
  \aap, 597, A81

\bibitem[{{Johnston} {et~al.}(2017b){Johnston}, {Hood}, {Cargill}, \& {De
  Moortel}}]{paper:Johnston2017b}
{Johnston}, C.~D., {Hood}, A.~W., {Cargill}, P.~J., \& {De Moortel}, I. 2017b,
  \aap, 605, A8

\bibitem[{{Klimchuk}(2006)}]{paper:Klimchuk2006}
{Klimchuk}, J.~A. 2006, \solphys, 234, 41

\bibitem[{{Klimchuk} {et~al.}(2008){Klimchuk}, {Patsourakos}, \&
  {Cargill}}]{paper:Klimchuk2008}
{Klimchuk}, J.~A., {Patsourakos}, S., \& {Cargill}, P.~J. 2008, \apj, 682, 1351

\bibitem[{{Laming}(2017)}]{Laming2017}
{Laming}, J.~M. 2017, \apj, 844, 153

\bibitem[{{Lionello} {et~al.}(2009){Lionello}, {Linker}, \&
  {Miki{\'c}}}]{paper:Lionello2009}
{Lionello}, R., {Linker}, J.~A., \& {Miki{\'c}}, Z. 2009, \apj, 690, 902

\bibitem[{{Mathioudakis} {et~al.}(2013){Mathioudakis}, {Jess}, \&
  {Erd{\'e}lyi}}]{Mathi2013}
{Mathioudakis}, M., {Jess}, D.~B., \& {Erd{\'e}lyi}, R. 2013, \ssr, 175, 1

\bibitem[{{McIntosh} {et~al.}(2012){McIntosh}, {Tian}, {Sechler}, \& {De
  Pontieu}}]{McIntosh2012}
{McIntosh}, S.~W., {Tian}, H., {Sechler}, M., \& {De Pontieu}, B. 2012, \apj,
  749, 60

\bibitem[{{McLaughlin} {et~al.}(2011){McLaughlin}, {de Moortel}, \&
  {Hood}}]{paper:McLaughlin2011}
{McLaughlin}, J.~A., {de Moortel}, I., \& {Hood}, A.~W. 2011, \aap, 527, A149

\bibitem[{{Miki{\'c}} {et~al.}(2013){Miki{\'c}}, {Lionello}, {Mok}, {Linker},
  \& {Winebarger}}]{paper:Mikic2013}
{Miki{\'c}}, Z., {Lionello}, R., {Mok}, Y., {Linker}, J.~A., \& {Winebarger},
  A.~R. 2013, \apj, 773, 94

\bibitem[{{Nakariakov} \& {Verwichte}(2005)}]{Nakariakov2005}
{Nakariakov}, V.~M. \& {Verwichte}, E. 2005, Living Reviews in Solar Physics,
  2, 3

\bibitem[{{Ofman} {et~al.}(1998){Ofman}, {Klimchuk}, \& {Davila}}]{Ofman1998}
{Ofman}, L., {Klimchuk}, J.~A., \& {Davila}, J.~M. 1998, \apj, 493, 474

\bibitem[{{Parnell} \& {De Moortel}(2012)}]{paper:ParnellDeMoortel2012}
{Parnell}, C.~E. \& {De Moortel}, I. 2012, Philosophical Transactions of the
  Royal Society of London Series A, 370, 3217

\bibitem[{{Prokopyszyn} {et~al.}(2019){Prokopyszyn}, {Hood}, \& {De
  Moortel}}]{Prokopyszyn2019}
{Prokopyszyn}, A.~P.~K., {Hood}, A.~W., \& {De Moortel}, I. 2019, \aap, 624,
  A90

\bibitem[{{Reale}(2010)}]{paper:Reale2010}
{Reale}, F. 2010, Living Reviews in Solar Physics, 7, 5

\bibitem[{{Reale}(2014)}]{Reale2014}
{Reale}, F. 2014, Living Reviews in Solar Physics, 11, 4

\bibitem[{{Terradas} \& {Ofman}(2004)}]{paper:Terradas2004}
{Terradas}, J. \& {Ofman}, L. 2004, \apj, 610, 523

\bibitem[{{Tripathi} {et~al.}(2012{\natexlab{a}}){Tripathi}, {Mason}, {Del
  Zanna}, \& {Bradshaw}}]{Tripathi2012}
{Tripathi}, D., {Mason}, H.~E., {Del Zanna}, G., \& {Bradshaw}, S.
  2012{\natexlab{a}}, \apjl, 754, L4

\bibitem[{{Tripathi} {et~al.}(2012{\natexlab{b}}){Tripathi}, {Mason}, \&
  {Klimchuk}}]{TripathiMason2012}
{Tripathi}, D., {Mason}, H.~E., \& {Klimchuk}, J.~A. 2012{\natexlab{b}}, \apj,
  753, 37

\bibitem[{{Verwichte} {et~al.}(2017){Verwichte}, {Antolin}, {Rowlands},
  {Kohutova}, \& {Neukirch}}]{Verwichte2017}
{Verwichte}, E., {Antolin}, P., {Rowlands}, G., {Kohutova}, P., \& {Neukirch},
  T. 2017, \aap, 598, A57

\bibitem[{{Verwichte} {et~al.}(1999){Verwichte}, {Nakariakov}, \&
  {Longbottom}}]{paper:Verwichte1999}
{Verwichte}, E., {Nakariakov}, V.~M., \& {Longbottom}, A.~W. 1999, Journal of
  Plasma Physics, 62, 219

\bibitem[{{Winebarger} {et~al.}(2013){Winebarger}, {Tripathi}, {Mason}, \& {Del
  Zanna}}]{Winebarger2013}
{Winebarger}, A., {Tripathi}, D., {Mason}, H.~E., \& {Del Zanna}, G. 2013,
  \apj, 767, 107

\bibitem[{{Withbroe} \& {Noyes}(1977)}]{paper:Withbroe1977}
{Withbroe}, G.~L. \& {Noyes}, R.~W. 1977, \araa, 15, 363

\bibitem[{{Zacharias} {et~al.}(2011){Zacharias}, {Peter}, \&
  {Bingert}}]{Zacharias2011}
{Zacharias}, P., {Peter}, H., \& {Bingert}, S. 2011, \aap, 532, A112

\bibitem[{{Zaqarashvili} \& {Erd{\'e}lyi}(2009)}]{Zaq2009}
{Zaqarashvili}, T.~V. \& {Erd{\'e}lyi}, R. 2009, \ssr, 149, 355

\end{thebibliography}

\appendix

\section{Ponderomotive Upflows} \label{appendix_pond_force}

\begin{figure*}[!t]
\centering
\includegraphics[width=0.49\textwidth]{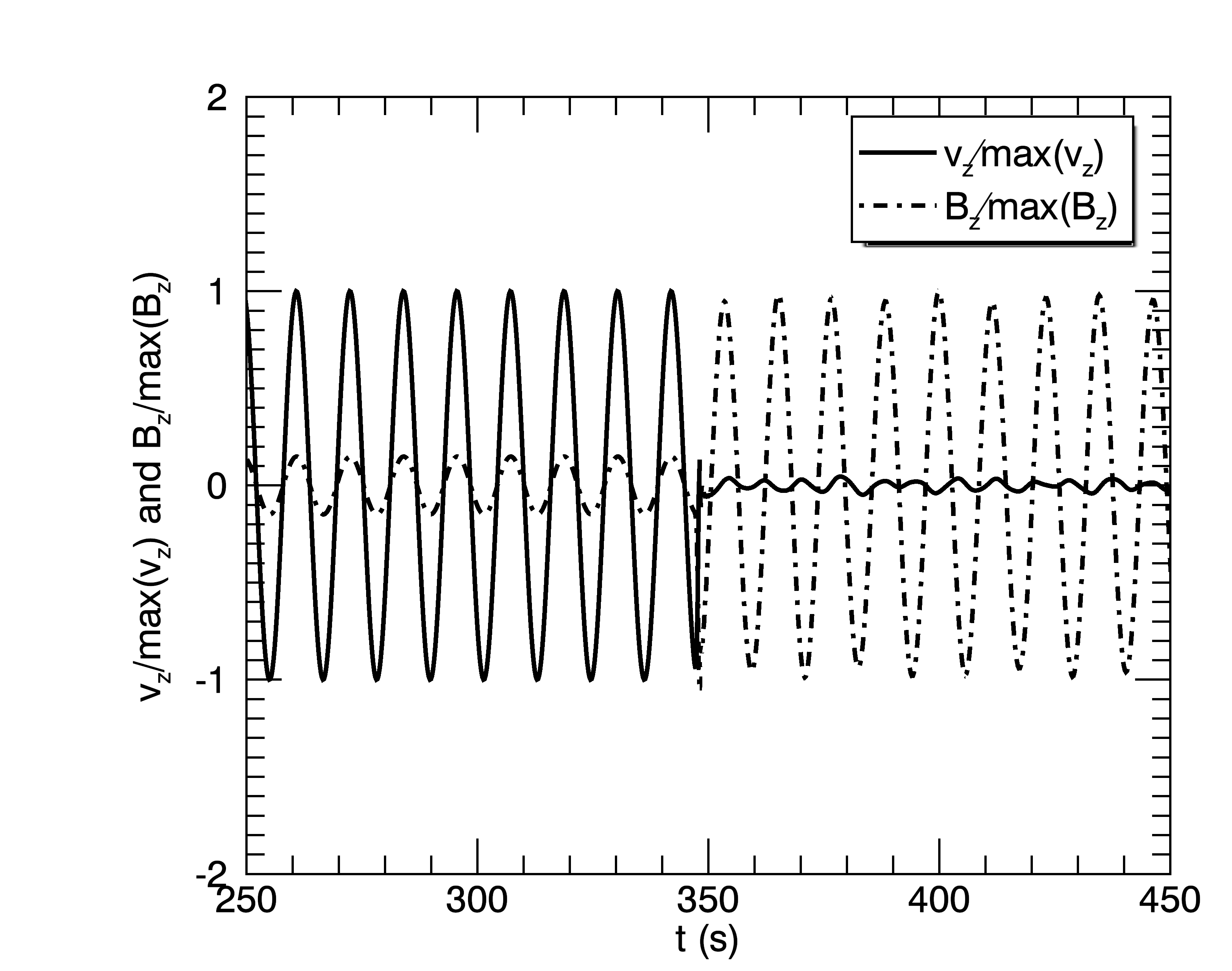}
\includegraphics[width=0.49\textwidth]{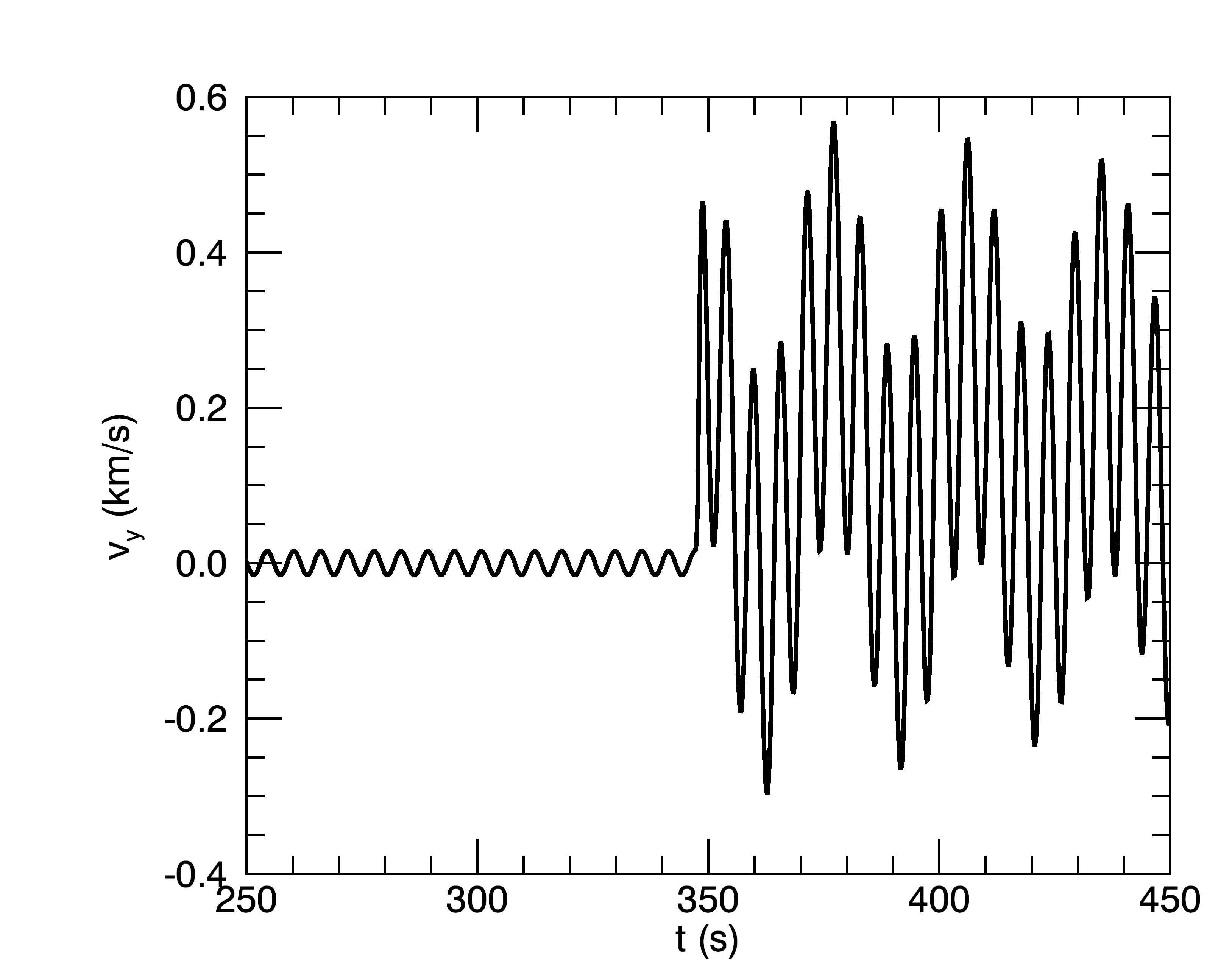}
   \caption{(Left) Plot of $v_z/\max(v_z)$ and $B_z/\max(B_z)$ with time at the location of the driver ($y=200$ Mm). (Right) Plot of the upflows $v_y$ (km/s) as a function of time at the location of the driver.}
\label{node_driver}
\end{figure*} 

In this appendix, we provide a more detailed description of the upflows associated with the Alfv\'en wave driver. Ponderomotive upflows, resulting from the non-linear magnetic pressure force \citep[e.g.][]{paper:Verwichte1999} have been discussed in the context of standing Alfv\'en waves in coronal loops \citep[see e.g.][]{paper:Terradas2004, paper:McLaughlin2011, Verwichte2017}. In our model setup however, the Alfv\'en waves are propagating, driven by an additional forcing term in the momentum equation (see Eqn.~\ref{driver}).

To isolate the effects of the ponderomotive force, we consider a simple 1D, uniform model and use the same force driver as in Eqn.~(\ref{driver}), situated at an arbitrary distance of $y=200$ Mm from the bottom boundary. This driver generates both left and right propagating Alfv\'en waves. After some time, the left-propagating waves reach the lower boundary (at $y=0$ Mm) where they reflect. We find that as these reflected waves (now right-propagating) interact with the left-propagating waves generated by the driver (at $y=200$ Mm), a standing mode is generated. After $t=347$s, the right-propagating waves have reached the location of the driver again, setting up a standing mode in the entire lower part of the domain ($0 < y < 200$ Mm). Hence, although the driver generates propagating waves, the reflection from the lower boundary indirectly leads to a standing wave in the lower part of the domain as the length of the lower part of the domain (i.e.~the distance between the bottom boundary and the driver) was chosen to be a multiple of the wavelength of the driven waves. Figure \ref{node_driver} (left) shows both the velocity (solid line) and magnetic (dot-dashed line) perturbations at the location of the driver ($y=200$ Mm) with time. We see that when the reflected waves reach the driver location (at $t=347$s), this location becomes a node in $v_z$ and an anti-node in $b_z$ from that time onwards. The presence of this standing wave node leads to a rapid increase in the magnetic pressure force, driving an upflow ($v_y$) as can be seen in Figure \ref{node_driver} (right). This upflow then leads to a compression of plasma, which propagates upwards at the slow speed.

\begin{figure}[h]
\centering
\includegraphics[width=0.49\textwidth]{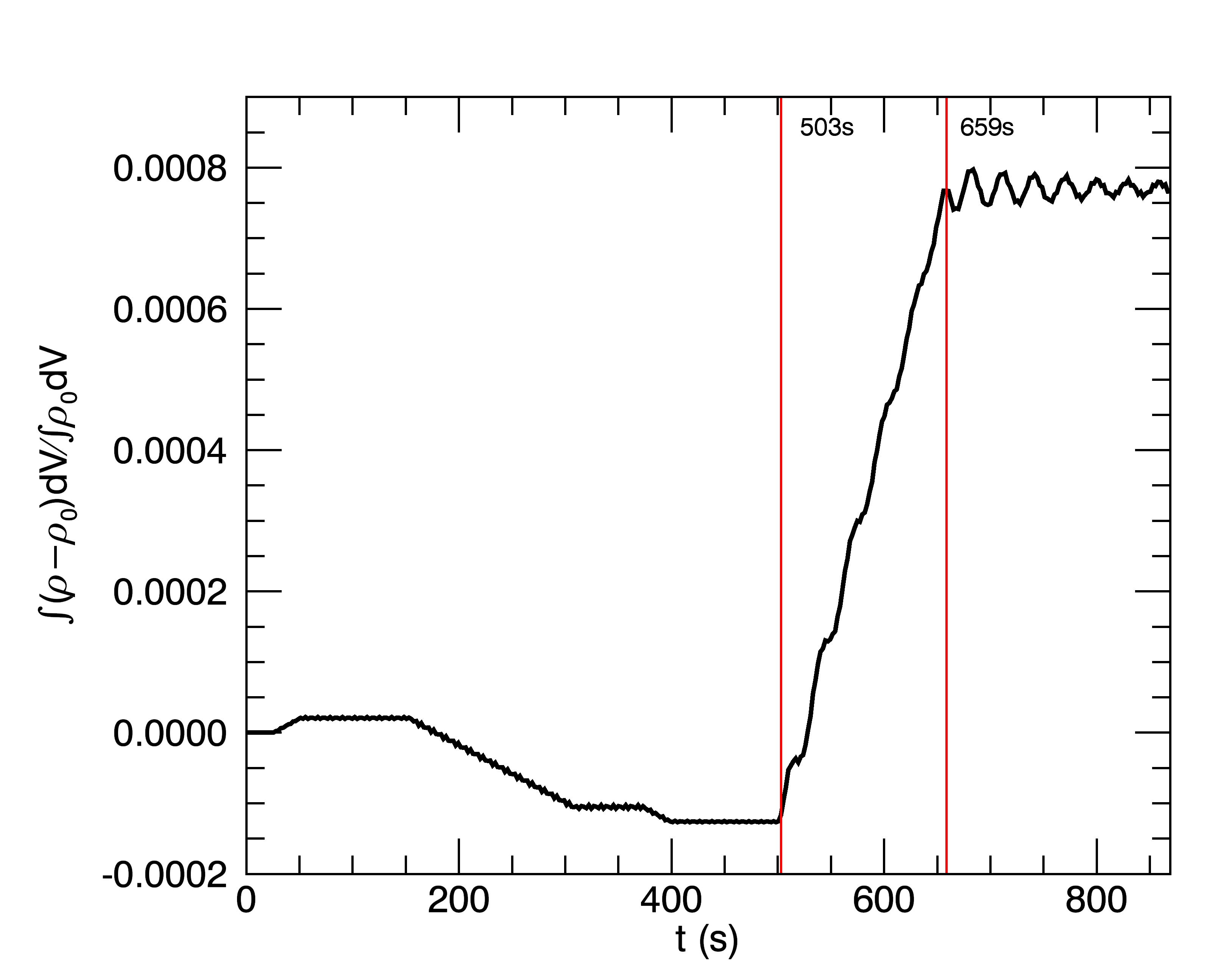}
   \caption{Plot of the relative mass increase $\int (\rho-\rho_0)dV/\int\rho_0 dV$ with time in the region $y=230-260$ Mm. The red vertical lines mark the times when the slow wave, that is generated at $t=347$s at the location of the driver, arrives and leaves this region.}
  \label{mass_increase_pond_1}
\end{figure}

Figure \ref{mass_increase_pond_1} shows the relative mass integrated over the region $y=230-260$ Mm, as a function of time. We can see a rapid increase in mass after $t=503$s, which accounts for the time required for the mass increase to propagate (at the slow speed) from the location of the driver (at $y=200$ Mm) to the region where we are measuring the mass increase (starting at $y=230$ Mm). By $t=659$s, the mass increase has reached $y=260$ Mm, which marks the end of the region in which we are sampling the mass increase and from then on, the mass stays constant in this interval.

\begin{figure}[h]
\centering
\includegraphics[width=0.49\textwidth]{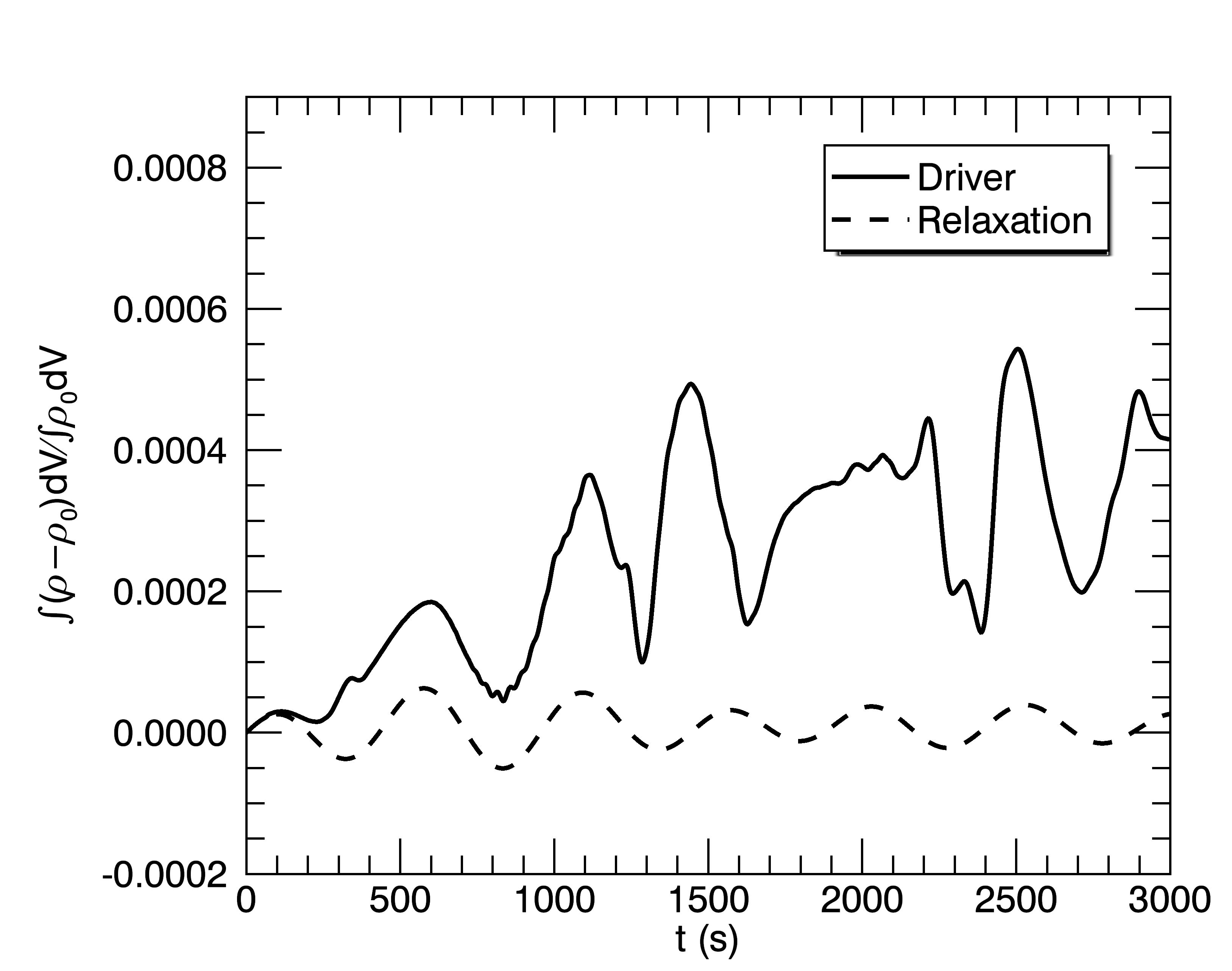}
   \caption{Plot of the relative mass increase $\int (\rho-\rho_0)dV/\int\rho_0 dV$ with time in the region $y=44-50$ Mm for the simulation that includes an atmosphere. The dashed line corresponds to the non-driven simulation.}
\label{mass_increase_pond_atmos}
\end{figure}

In the example described above, the lower boundary was set to be a perfectly reflecting boundary. In the model used in Section \ref{s:Model}, the downward (left) propagating waves will reflect of the density gradient in the chromosphere but this reflection happens gradually rather than at a defined point (such as the lower boundary in the simple model described above). Figure \ref{mass_increase_pond_atmos} shows the relative mass increase for a 1D model with the same field-aligned thermodynamic equilibrium as our 2D model described in Section \ref{s:Model}. As in the 2D model, the driver is now located at $y=7.8$ Mm (the top of the first chromosphere). The region where the mass is evaluated is taken as $y=44-50$ Mm (in the corona). As the reflection of the downward propagating waves is now happening gradually rather than instantaneously, the mass increase (solid line) is happening more gradually as well.

\end{document}